\documentclass[
    reprint,
    aps,
    pra,
    twocolumn,
    amsmath,
    amssymb,
    nofootinbib,
]{revtex4-1}

\usepackage{graphicx}      
\usepackage{dcolumn}       
\usepackage{bm}            
\usepackage{microtype}     
\usepackage{braket}
\usepackage{algorithm}
\usepackage{algpseudocode}
\usepackage{hyperref}

\newcommand*{\logneg}{\ensuremath{\mathcal{E}}}
\newcommand*{\rhopt}{\ensuremath{\rho_{AB}^{T_B}}}
\newcommand*{\tr}{\ensuremath{\text{Tr}}}

\begin{document}

\title{Fast Computation of Many-Body Entanglement}
\author{Johnnie Gray}
\email{john.gray.14@ucl.ac.uk}
\affiliation{University College London}

\begin{abstract}
Mixed state entanglement measures can act as a versatile probes of many-body systems.
However, they are generally hard to compute, often relying on tricky optimizations.
One measure that is straightforward to compute is the logarithmic negativity, yet done naively even this is still limited to small system sizes.
Here, we introduce a method to compute the logarithmic negativity for arbitrary subsystems of a densely represented state, as well as block subsystems of matrix product states.
The method combines lazily evaluated, tensor network representations of the partially transposed density matrix with stochastic Lanczos quadrature, and is easily extendible to other quantities and classes of many-body states.
As examples, we compute the entanglement within random pure states for density matrices of up to 30 qubits,  explore scrambling in a many-body quench, and match the results of conformal field theory in the ground-state of the Heisenberg model for density matrices of up to 1000 spins.
An implementation of the algorithm has been made available in the open-source library \textit{quimb}.
\end{abstract}

\maketitle

\section{Introduction}
Entanglement not only plays an essential role across many aspects of quantum technologies~\cite{rao1945information,shor1999polynomial,harrow2009quantum,bennett1993teleporting,bennett1992communication,ekert1991quantum}, but also in understanding the nature of many-body quantum systems~\cite{amico2008entanglement,schollwock2011density}.
A prevalent quantity to study in this context is the \emph{entanglement entropy}, either computationally or analytically~\cite{calabrese2012entanglement,nishioka2009holographic,calabrese2009entanglement}.
However, this is only applicable to bipartitions of pure states: one can only control the ratio of subsystem sizes and cannot, for example, exclude any sort of environment.
On the other hand, a true mixed-state entanglement measure allows full control over the sizes of two subsystems at once, and can thus be a much more refined probe for many phenomena~\cite{wichterich2009scaling,bayat2010negativity,caruso2010entanglement,calabrese2012entanglement,calabrese2013entanglement,eisler2014entanglement,wen2015entanglement,sherman2016nonzero,bayat2017scaling,gray2018many}.
One drawback is that most true entanglement measures are inefficient to compute for many-body systems, even ignoring the exponential scaling of Hilbert space size, $d$, with system size, $L$.
One quantity that is efficient~\cite{huang2014computing}, in a technical sense, is the \emph{logarithmic negativity}~\cite{zyczkowski1998volume,lee2000partial,vidal2002computable,plenio2005logarithmic}, though the naive computational effort still scales cubically with Hilbert space size, limiting practical calculations to $\lesssim 15$ qubits.

Here we demonstrate an efficient method to approximately but accurately compute the logarithmic negativity for subsystems of many-body quantum states.
The method relies on treating the reduced density matrix as an implicit operator defined as a tensor network~\cite{schollwock2011density,orus2014practical,bridgeman2017hand}, and then using stochastic Lanczos quadrature~\cite{lanczos1950iteration,golub1994matrices,ubaru2017fast} to estimate a spectral sum of this operator.
We refer to the whole procedure as tensor network stochastic Lanczos quadrature (TNSLQ).
The logarithmic negativity is a particular instance of the algorithm, which we target here, but other quantities such as entropy and thus mutual information are even simpler to compute.
There are also many representations of many-body states amenable to a tensor network description, but we focus here on two key ones: (i) density operators derived from partially tracing densely represented pure states; and (ii) `compressed' density operators derived from partially tracing matrix product states.
Broadly speaking, the TNSLQ method enables the computation of logarithmic negativity for density matrices of $\lesssim 30$ qubits, without resorting to supercomputer-level resources.
In terms of matrix product states, the equivalent limit for computing entanglement between arbitrarily separated contiguous blocks, with open or periodic boundary conditions, is that the the bond dimension is initially $\lesssim 180$.
Efficient implementations of the algorithm specifically for both of these classes of states have been added to the open source library \texttt{quimb}~\cite{gray2018quimb}, as well as the general capability to perform TNSLQ for arbitrary tensor networks and quantities.

This paper is organised as follows: in Sec.~\ref{sec:logneg} we introduce the logarithmic negativity and discuss some details of its naive computation.
In Sec.~\ref{sec:SLQ} we introduce stochastic Lanczos quadrature as a method to approximately compute the logarithmic negativity as the spectral sum of a linear operator.
In Sec.~\ref{sec:tn-intro} we introduce the basic diagrammatic notation of tensor networks.
In Sec.~\ref{sec:ptr-states} we show how to form an efficient partially transposed linear operator for two subsystems of an exactly represented pure state.
In Sec.~\ref{sec:mps} we show how to do the same for two block subsystems of matrix product states, which involves a form of `compression' first.
In Sec.~\ref{sec:results} we present results of using the above methods as applied to relevant physical situations.
In Sec.~\ref{sec:error-analysis} we analyse the error of the method and show that it is bounded by the purity of the density operator under consideration.
Finally, we discuss the method's future applications and conclude in Sec.~\ref{sec:discussion}.

\section{Logarithmic Negativity \label{sec:logneg}}

The logarithmic negativity~\cite{zyczkowski1998volume,lee2000partial,vidal2002computable,plenio2005logarithmic} is an entanglement monotone and upper bound on the distillable entanglement.
For a density matrix, $\rho_{AB}$, of two subsystems $A$ and $B$ with Hilbert space sizes $d_A$ and $d_B$ respectively, it is defined as
\begin{equation} \label{eq:logneg}
    \logneg(\rho_{AB}) = \log_2 \left\|\rhopt \right\|_\tr
\end{equation}
with $\cdot^{T_B}$ denoting the \emph{partial transpose}~\cite{peres1996separability} with respect to subsystem $B$ and $\|\cdot\|_\tr$ the \emph{trace norm}\footnote{Also known as the \emph{nuclear norm}.}.
Unlike the mutual information say, the logarithmic negativity quantifies quantum correlations only - one of the features that mark it out as a refined probe of many-body quantum phenomena.

We note that even if $\rho_{AB}$ is a low-rank operator, the partial transpose operation generally increases the rank by a factor of $\min\{ d_A^2, d_B^2 \}$, precluding the use of low-rank methods for the computation of \logneg.
Instead, the trace norm of an operator is generally computed as the absolute sum of all eigenvalues, and as such, the full spectrum is required in the exact case.
On the other hand, if much of the spectrum can be essentially described as a continuous distribution, then intuition suggests that far less information than every single eigenvalue should be required to approximate its sum.
In this case it should also be possible to avoid directly forming the full, partially transposed, density operator \rhopt~and instead rely only on its \emph{action} on an arbitrary vector: $\rhopt \ket{\phi} \rightarrow \ket{\tilde \phi}$.
We'll call such an implicit representation simply a \emph{linear operator}, $\hat X$.

\section{Stochastic Lanczos Quadrature \label{sec:SLQ}}

Let's assume we have access to \rhopt~as a linear operator, that is, we can use it to evaluate matrix-vector products.
We can also recast Eq.~\eqref{eq:logneg} as the trace of a matrix function where we take the function as the \emph{absolute} function, $|\cdot|$:
\begin{equation}\label{eq:logneg-as-trace}
    \logneg(\rho_{AB}) = \log_2 \tr \left(  \left| \rhopt \right| \right)~.
\end{equation}
For such a \emph{spectral sum} of a Hermitian linear operator there do indeed exist various methods to estimate the quantity, including polynomial methods~\cite{han2016approximating} and approximate reconstruction of the spectrum~\cite{lin2016approximating}
We focus here though on \emph{Stochastic Lanczos Quadrature} (SLQ)~\cite{ubaru2017fast}, which is relatively simple to implement but also exhibits excellent performance.
It can be thought of as the combination of three separate techniques:
\begin{enumerate}
    \item
    \textbf{Hutchinson's trace method}\cite{hutchinson1990stochastic}, which estimates the trace of an operator, $f(\hat X)$, with $N$ inner product samples of random vectors $\{\ket{\phi_n}\}$:
    \begin{equation}
        \tr (f(\hat X)) \approx \frac{1}{N} \sum_{i=1}^{N} \Braket{\phi_n | f(\hat X) | \phi_n}~.
    \end{equation}
    This approaches the exact value, $\tr (f(\hat X))$, in a unbiased manner as $N\rightarrow\infty$.
    Practically speaking, we generally need $N \ll d_A d_B$ in order to estimate the trace to reasonable accuracy.
    For the purpose of the logarithmic negativity we take $f\rightarrow |\cdot|$  and $\hat X \rightarrow  \rhopt$ so that $f(\hat X) \rightarrow |\rhopt|$.
    \item
    \textbf{Gauss Quadrature}, which allows the estimation of the above bi-linear forms, $G_n = \Braket{\phi_n | f(\hat X) | \phi_n}$, when transformed into a Riemann--–Stieltjes integral~\cite{ubaru2017fast}.
    Note that the vector $\ket{\tilde y} = f(\hat X)\ket{\phi_n}$ is not itself directly computed at any point, which would be expensive.
    \item
    The \textbf{Lanczos algorithm}~\cite{lanczos1950iteration}, which iteratively constructs a basis for the Krylov space $\textrm{span}\{\ket{\phi_n}, \hat X^1 \ket{\phi_n}, \hat X^2 \ket{\phi_n}, \ldots\}$ using matrix-vector products only, from which the nodes and weights of the Gauss quadrature rule can be directly computed.
\end{enumerate}
Details of each of these three techniques, including error analysis, are extensively addressed in various other publications~\cite{golub1994matrices,ubaru2017fast}.
Instead, we simply sketch a full implementation of the SLQ method in Algorithm~\ref{alg:SLQ}.

\begin{algorithm}[H]
  \caption{
  \emph{Inputs:}
  linear operator $\hat X$,
  scalar function $f$,
  target accuracy \texttt{tol},
  target Lanczos accuracy \texttt{ltol},
  maximum number of repeats $N_\text{max}$,
  maximum Krylov subspace size $K_\text{max}$.
  \emph{Outputs:} SLQ estimate of $\tr(f(\hat X))$
  }
  \label{alg:SLQ}
   \begin{algorithmic}[1]
   \For{$n = 1:N_\text{max}$}
      \State{$\ket{\phi} \gets$ zero mean, unit variance, random vector}
      \State{$\beta_1 = \sqrt{\braket{\phi | \phi}}$}
      \State{$\ket{\phi_0} \gets 0$}
      \State{$\ket{\phi_1} \gets \ket{\phi} / \beta_1$}
      \For{$k = 1:K_\text{max}$}
          \State{// \textit{perform a Lanczos iteration}}
          \State{$ \ket{\tilde{v}} = \hat X \ket{\phi_k} - \beta_k \ket{\phi_{k - 1}}$}
          \State{$ \alpha_k = \braket{\tilde{v} | \phi_k} $}
          \State{$ \ket{\tilde{v}} = \ket{\tilde{v}} - \alpha_k \ket{\phi_k}$}
          \State{$ \beta_{k + 1} = \sqrt{\braket{\tilde{v} | \tilde{v}}} $}
          \State{$ \ket{\phi_{k - 1}} \gets \ket{\phi_{k}} $}
          \State{$ \ket{\phi_{k}} = \ket{\tilde{v}} / \beta_{k + 1} $}
          \State{// \textit{compute Gauss quadrature weights}}
          \State{$ T_k = \textbf{tridiag}([ \alpha_1, \ldots, \alpha_k ], [ \beta_2, \ldots, \beta_k ]) $}
          \State{$ \{ \theta_j\}, \{ \ket{w_j} \} = \textbf{eig}(T_k)$}
          \State{compute $ \tau_j = \braket{e_1 | w_j}$ for each eigenvector $\ket{w_j}$}
          \State{$ \tilde{F}_k = \sum_{j} \tau_j^2 f(\theta_j) $}
          \If{$
          \texttt{lanczos\_converged}([\tilde{F}_1, \ldots,\tilde{F}_{k}],
          \texttt{ltol}) $}
            \State{\textbf{break}}
          \EndIf
      \EndFor
      \State{$G_n \gets \texttt{lanczos\_estimate}([\tilde{F}_1, \ldots,\tilde{F}_{k}])$}
      \If{\texttt{hutchinson\_converged($[G_{1}, \ldots, G_n]$, \texttt{tol})}}
        \State{\textbf{break}}
      \EndIf
   \EndFor
   \State{\textbf{return} $1 / n \sum_{j=1}^{n} G_j$}
   \end{algorithmic}
\end{algorithm}

We note that the SLQ algorithm requires storage of 3 vectors of size $d_A d_B$ only -- a relatively low memory overhead.
For large systems, the computational effort is generally dominated by the matrix-vector product $\hat X \ket{\phi_k}$.
There are multiple options for choosing the Lanczos convergence and estimation functions \texttt{lanczos\_converged} and \texttt{lanczos\_estimate}~\cite{bellalij2015bounding}.
We find a practical method is to least-squares fit an exponential to the values $[\tilde{F}_1, \ldots,\tilde{F}_{k}]$.
This yields a value for the equilibrium point, as well as an uncertainty, which can respectively be taken as an estimate for the bilinear form and an error to test convergence against.
For \texttt{hutchinson\_converged} we simply take the error on the mean.
The overall procedure not only allows us to estimate quantities such as the logarithmic negativity, but reliably keep track of the error as well.

\section{Tensor Networks \& Graphical Notation \label{sec:tn-intro}}

\begin{figure}[tb]
    \centering
    \includegraphics[width=\linewidth]{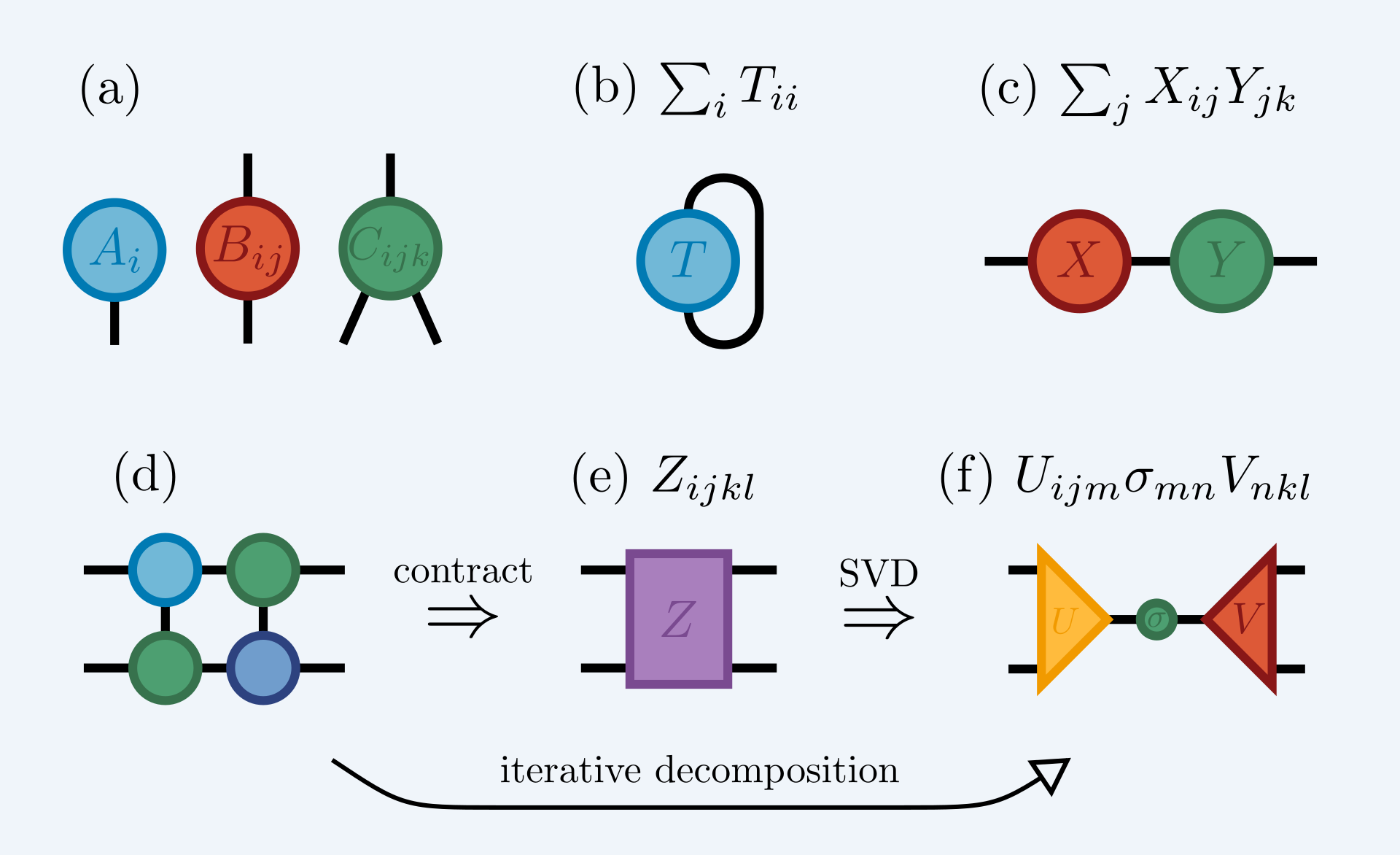}
    \caption{\emph{Tensor diagrams:} (a) a vector, matrix and rank-3 tensor in graphical form; (b) the trace of a matrix; (c) tensor contraction, here matrix multiplication; (d) a small network of tensors, which could be contracted into a single tensor (e), itself then decomposed, for example via SVD, into a new network (f).
    The left and right indices of (d) could also be used to treat it as a linear operator.
    For here it could be iteratively decomposed, directly into (f).}
    \label{fig:tensor-network-basics}
\end{figure}

The remaining task is to find a linear operator representation of \rhopt~ for our target class of many-body states.
We will focus here on using a \emph{tensor network} to represent this implicit operator and perform matrix-vector products.
As such we'll first briefly recap the graphical notation associated with tensor networks --  more thorough reviews can be found in~\cite{schollwock2011density,orus2014practical,bridgeman2017hand,biamonte2017tensor}.
The essential idea is to treat all quantum objects as \emph{tensors}, i.e. n-dimensional objects describing linear mappings between spaces, with a labelled index for each dimension.
For the purpose of finite quantum mechanics, these tensors are simply numeric arrays, $T_{ij\ldots k}$

The basic graphical notation is shown in Fig.~\ref{fig:tensor-network-basics}.
We depict tensors as shapes/nodes, with a leg/edge representing each index.
Scalars thus have no legs, vectors one leg, and matrices two legs.
A $L$-body pure quantum state, $\ket{\psi}$, we can view as a rank-$L$ tensor\footnote{Here we mean `rank' as the number of indices, or dimensions, of the tensor, rather than number of non-zero singular values.}, $\psi_{1\ldots L}$.
Connecting the legs of tensors implies a combined summation over that shared index -- a contraction, see Figs.~\ref{fig:tensor-network-basics}(b), (c).
In this way networks of tensors can be built up, with the number of free legs indicating the rank of the full, lazily represented object (see Fig.~\ref{fig:tensor-network-basics}(d) $\Rightarrow$ (e)).
If evaluating a tensor network, it is always most efficient to perform a series of pairwise contractions, the order of which can massively affect performance.
Indices can be arbitrarily grouped into new, larger indices (or if their dimension factorizes, ungrouped).
This `vectorization' allows any tensor contraction to be performed as either a vector-vector, matrix-vector or matrix-matrix product.
Tensors can also be decomposed, for example via singular value decomposition (SVD), into a new tensor network -- see Fig.~\ref{fig:tensor-network-basics}(e) $\Rightarrow (f)$.

For any network, or sub-network, we can also mark the open indices as either `left' or `right' and treat the resulting object as a linear operator which maps vectorized tensors spanning one set of indices into the other.
The key here is that if only the action of such a linear operator on a vector is required, then the full operator does not need to be formed, and instead, the vector can be efficiently contracted into the tensor network.
This allows iterative decompositions that directly transform Fig.~\ref{fig:tensor-network-basics}(d) into Fig.~\ref{fig:tensor-network-basics}(f), for example.
One such useful procedure is the interpolative SVD~\cite{liberty2007randomized,woolfe2008fast,martinsson2011randomized}, which can be used to estimate the rank of the lazily represented operator to a certain precision, and then perform the decomposition to that target rank.
And another possible procedure is of course the SLQ algorithm described above.

The TNSLQ method is thus to take a tensor network, form a lazily represented linear operator, $X$, by grouping indices into `left' or `right' sets, then perform SLQ using the fact the sampling vector (which is really a vectorized tensor) can efficiently be contracted \emph{into} the network to estimate quantities of the form $\tr{f(X)}$.
We note that in general, such operators do not have a sparse-matrix linear operator representation, and might also be \emph{full-rank}, in the sense that all their singular values are significant.
Nonetheless, the TNSLQ method is applicable.

\section{Partial Trace States \label{sec:ptr-states}}

\begin{figure}[tb]
    \centering
    \includegraphics[width=\linewidth]{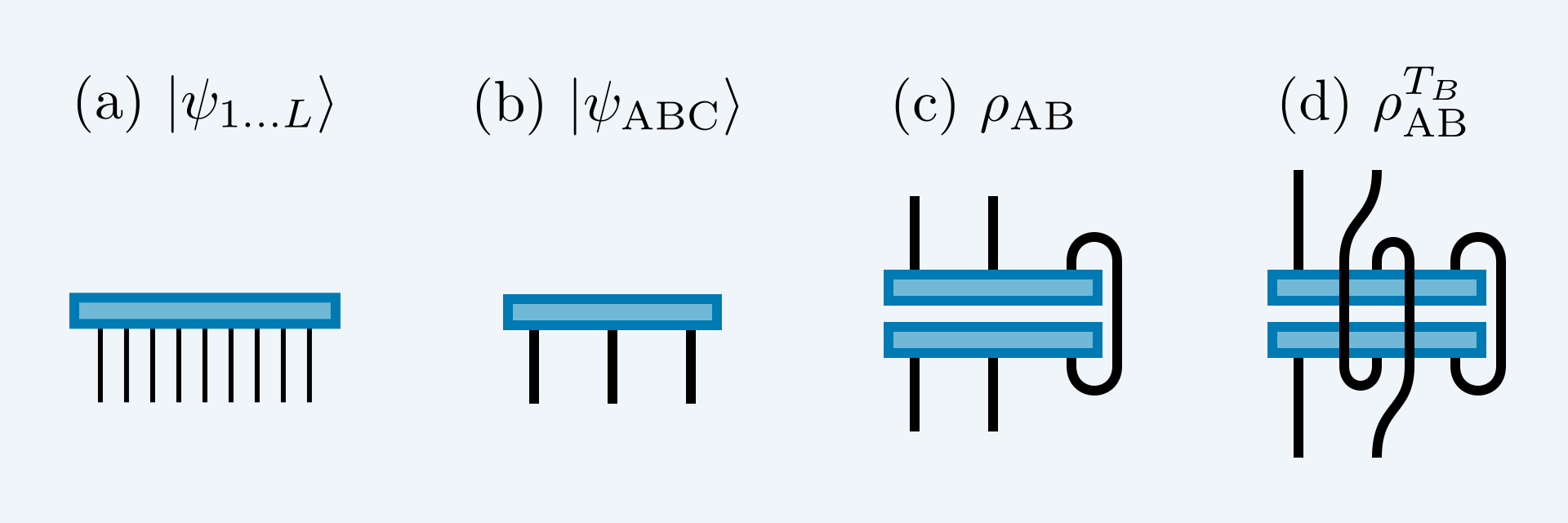}
    \caption{Steps to form a tensor network linear operator representation of \rhopt:
    (a) $\rightarrow$ (b) -- group indices into subsystems $A$, $B$ or $C$;
    (b) $\rightarrow$ (c) -- form the vector outer product and trace out system $C$;
    (c) $\rightarrow$ (d) -- exchange the `bra' and `ket' indices of subsystem $B$ to perform the partial transpose.
    For both (c) and (d) the linear operator representation of the operator is taken by grouping the upper and lower indices respectively.
    }
    \label{fig:exact-form-rhopt}
\end{figure}

\begin{figure}[tb]
    \centering
    \includegraphics[width=\linewidth]{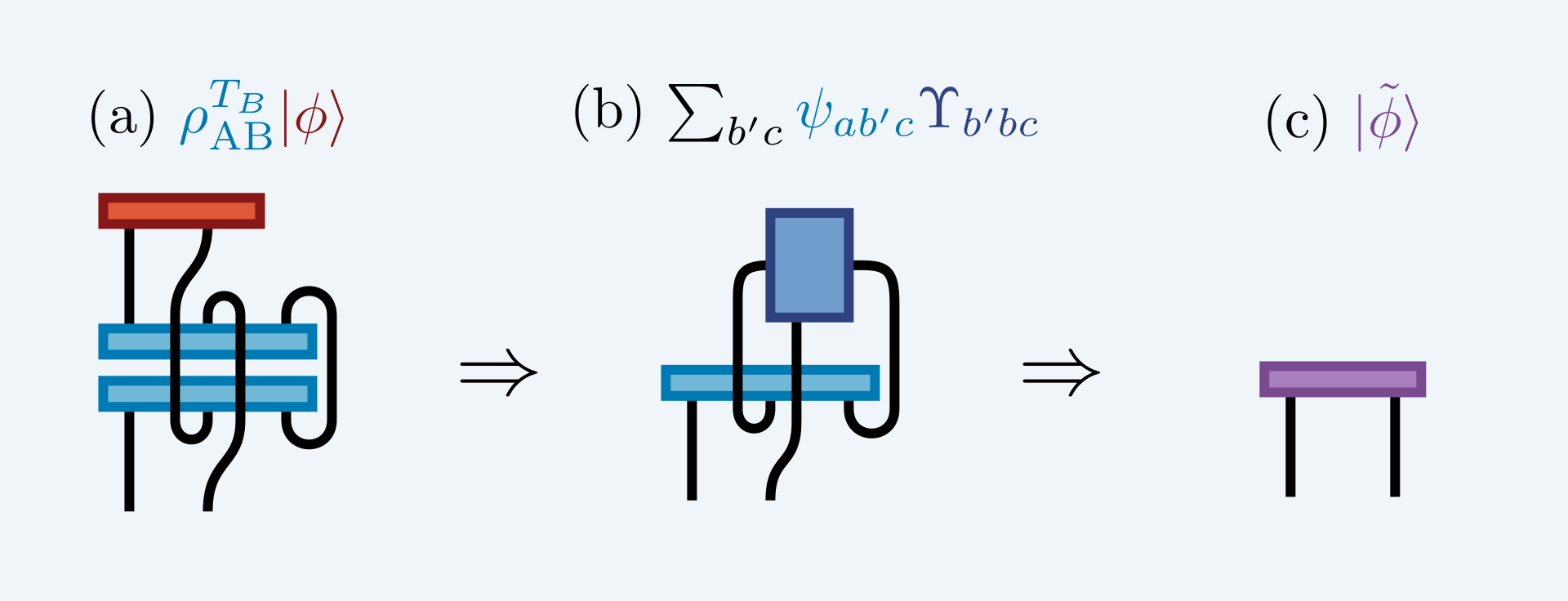}
    \caption{
    Acting on a vector with the linear operator $\rhopt$, formed as in Fig.~\ref{fig:exact-form-rhopt}: (a) the full tensor network describing $\rhopt \ket{\phi}$; (b) one of three possible intermediate contractions; and (c) the resulting output vector of this contraction.
    Note that although the tensor network in (a) \emph{represents} a vector, it can not be manipulated as such until it is contracted down to the form (c), the efficiency of which drastically depends on which intermediaries are chosen.
    }
    \label{fig:exact-rhopt-contract}
\end{figure}

Having briefly introduced SLQ and tensor networks as linear operators, we now move onto specific instances of many-body quantum states with bipartite density matrix subsystems that can be described in this way.
The first such example we'll call \emph{partial trace states} (PTS).
These are not genuine tensor networks in the sense that there is no entanglement induced geometry in the initial state, but the graphical notation is useful nonetheless.
The starting point is an exactly represented pure state vector -- with no particular requirements on the subsystem structure.
Without loss of generality we can take the many-body case of a pure $L$-body wave-function $\ket{\psi_{1 \ldots L}}$ -- a rank-$L$ tensor (Fig.~\ref{fig:exact-form-rhopt}(a)).
By grouping indices into either subsystem $A$, $B$ or $C$ -- where we want to trace out $C$ then find the entanglement between $A$ and $B$ -- we get a rank-3 tensor, $\ket{\psi_{ABC}}$, of total size $d_A d_B d_C$ (Fig.~\ref{fig:exact-form-rhopt}(b)).
The next step is to form an outer product with the conjugated state and lazily trace out subsystem $C$ to form $\rho_{AB}$ (Fig.~\ref{fig:exact-form-rhopt}(c)).
Finally we partially transpose the operator by swapping the `bra' and `ket' indices of subsystem $B$ to form \rhopt.
The advantage of keeping this operator represented as a tensor network is that is that the total storage remains $\propto d_A d_B d_C$.
Whereas clearly any time that $d_C < d_A d_B$ actually performing the partial trace would increase memory usage, potentially drastically, to $d_A^2 d_B^2$.

Since we want to now perform the SLQ procedure on this lazily represented tensor network operator we need to inspect how to act with it on a vector, $\ket{\phi}$, spanning the Hilbert space of subsystems $A$ and $B$.
In standard tensor notation we have:
\begin{equation}
  \rhopt \ket{\phi} = \sum_{a',b',c} \psi_{ab'c} \psi^{*}_{a'bc} \phi_{a'b'} ~,
\end{equation}
for which there are three possible intermediaries:
(i) $\sum_{b', c} \psi_{ab'c} \Upsilon_{b'bc}$;
(ii) $\sum_{a', c} \tilde{\Upsilon}_{aa'c} \psi^{*}_{a'bc}$; and
(iii) $\sum_{a',b'} \rho_{ab'a'b} \phi_{a'b'}$; the last of which is equivalent to explicitly forming the partially traced, partially transposed density matrix.
The dimensions of the subsystems determine which intermediary is best to form - in Fig.~\ref{fig:exact-rhopt-contract} we demonstrate performing the full contraction using the first intermediary to yield the new vector $\ket{\phi}$.
Equipped with this lazy linear operator representation of the \rhopt, we can now apply the SLQ procedure as detailed in Algorithm.~\ref{alg:SLQ} to compute the logarithmic negativity according to Eq.~\ref{eq:logneg-as-trace}.

Clearly we are still limited by needing to explicitly represent the full pure state $\psi_{ABC}$ (to $\lesssim$ 30 qubits on a `standard' desktop computer). However, the need to explicitly represent the full operator \rhopt~is lifted, allowing the computation of entanglement for \emph{any} tri-partition of $A$, $B$ and $C$.
Take for example the scenario where $L=30$  and each subsystem consists of 10 qubits.
Forming \rhopt~ would require about 16 terabytes of memory, let alone the time to fully diagonalize it, with the situation becoming even more extreme as we decrease the size of $C$.
On the other hand, with this lazy TNSLQ method, it is an easily tractable computation without a super-computer.

\section{Matrix Product States \label{sec:mps}}

\begin{figure}[tb]
  \centering
  \includegraphics[width=\linewidth]{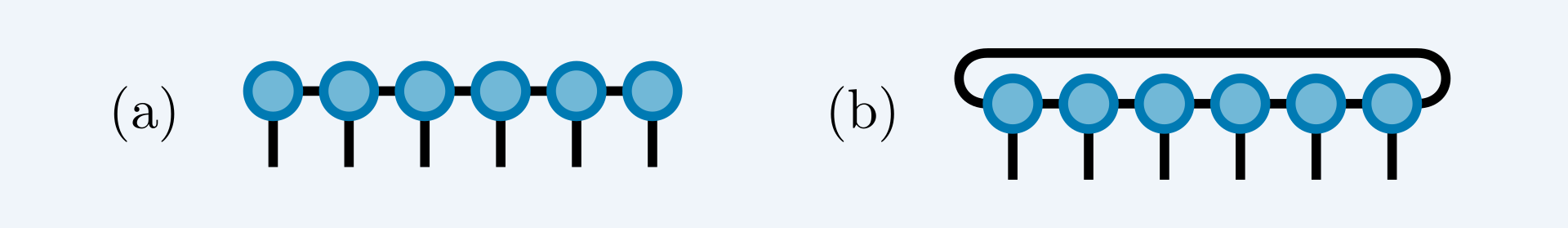}
  \caption{A $L=6$ MPS decomposition of a wavefunction with: (a) open boundary conditions, where the edge tensors are only rank-2; and (b) periodic boundary conditions.}
  \label{fig:mps-def}
\end{figure}

In order to move beyond full Hilbert space representations of many-body states we need a genuine tensor-network decomposition.
The most useful and widespread of these is that of the \emph{matrix product state} (MPS), which factorizes the wavefunction into a one-dimensional chain of rank-3 tensors.
This ansatz efficiently represents one-dimensional states with area-law entanglement~\cite{hastings2007area} and is the central representation in successful algorithms such as density matrix renormalization group (DMRG)~\cite{white1992density,schollwock2011density} and time evolving block decimation~\cite{vidal2003efficient}.
The form can be explicitly defined as
\begin{equation}\label{eq:mps-def}
  \psi_{abc\ldots z} =
  \sum_{\alpha,\beta,\gamma,\delta \ldots, \zeta}
  A_{a\alpha\beta}
  B_{b\beta\gamma}
  C_{c\gamma\delta}
  \ldots
  Z_{z\zeta\alpha}
\end{equation}
for tensors $A,B,C,\ldots,Z$ with physical indices $a,b,c,\ldots,z$, but it is generally more concise to reason with the graphical notation as depicted in Fig.~\ref{fig:mps-def}.
For simplicity we will consider the size of all the physical indices to be $p$, and the size of all the virtual indices, $\alpha,\beta,\gamma,\delta \ldots, \zeta$, (the \emph{bond dimension}) to be the same value, $\chi$.
The index $\alpha$ can be taken as size 1 (and thus ignored) for open boundary conditions - Fig.~\ref{fig:mps-def}(a) - or $\chi$ for periodic boundary conditions - Fig.~\ref{fig:mps-def}(b).

\begin{figure}[tb]
  \centering
  \includegraphics[width=\linewidth]{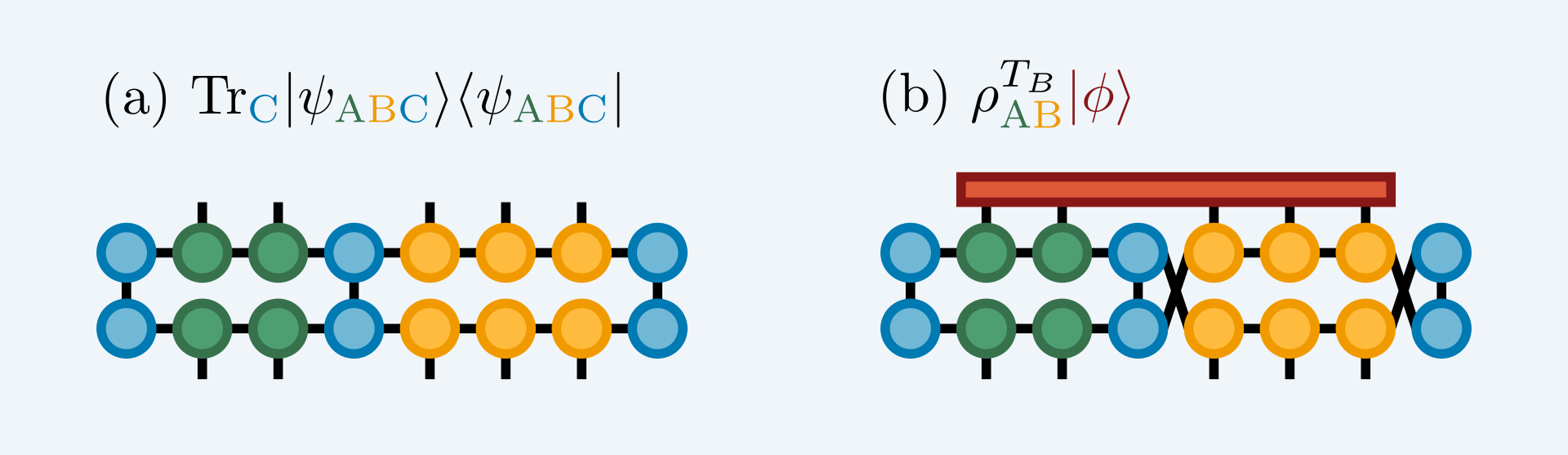}
  \caption{
  (a) Forming the reduced density matrix $\rho_\text{AB}$ from matrix product state $\psi_\text{ABC}$. Here subsystem $A$ is in green, subsystem $B$ is in yellow, and subsystem $C$ is in blue.
  (b) Tensor network representation of acting on a dense vector, $\ket{\phi}$ (in red), with \rhopt~ derived from the matrix product state $\psi_\text{ABC}$.
  }
  \label{fig:mps-rhoab-pt-lo}
\end{figure}

Given an MPS with target subsystems $A$ and $B$ to find the entanglement between, is is straightforward to form a tensor network of \rhopt. The steps as are follows:
(i) form the outer product between a `ket' and `bra' of the state;
(ii) perform the partial trace of environment $C$ by contracting (joining) all physical indices not contained in subsystems $A$ or $B$ (shown in Fig.~\ref{fig:mps-rhoab-pt-lo}(a));
and (iii) perform the partial transpose by switching the `ket' indices with the `bra' indices of all the physical sites in either subsystem $A$ or $B$.
At this point we could directly form a linear operator by grouping all the `ket' indices and `bra' indices respectively.
In this case, to perform the SLQ procedure we would then need to sample this operator using a dense vector of size $2^{L_{AB}}$ (for qubits), as shown in Fig.~\ref{fig:mps-rhoab-pt-lo}(b).
Here, we are now limited $L_{AB} \lesssim 30$ (rather than total length $L$ in the pure state subsystem case).

\begin{figure}[tb]
    \centering
    \includegraphics[width=\linewidth]{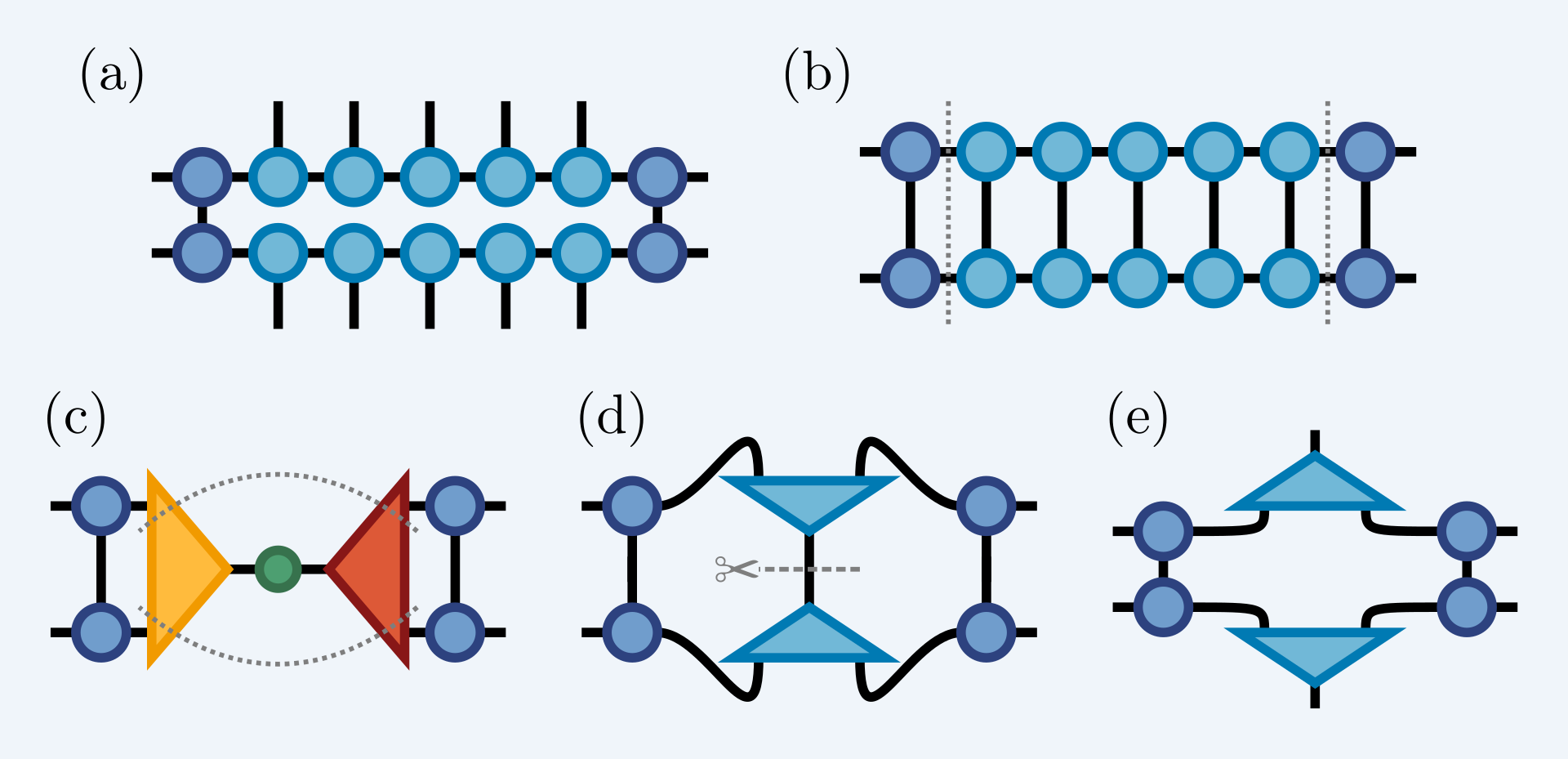}
    \caption{
    Compressing a MPS section in four steps.
    (a) $\Rightarrow$ (b): the overlap of the target section, or transfer matrix, is formed.
    (b) $\Rightarrow$ (c): this transfer matrix is laterally compressed using, for example, an iterative SVD algorithm (dotted grey lines in (b) denote the `left' and `right' groupings of indices).
    The singular values (central green tensor in (c)), can be absorbed into the left (yellow) or right (red) tensors once small values have been trimmed.
    (c) $\Rightarrow$ (d): this newly compressed section is vertically decomposed, for example using a Cholesky decomposition (dotted grey lines in (c) denote the new `left' and `right' groupings of indices).
    (d) $\Rightarrow$ (e): the bond between the new symmetric factors is `cut' in order to expose the new physical index.
    }
    \label{fig:mps-compress-section}
\end{figure}

\begin{figure}[tb]
    \centering
    \includegraphics[width=\linewidth]{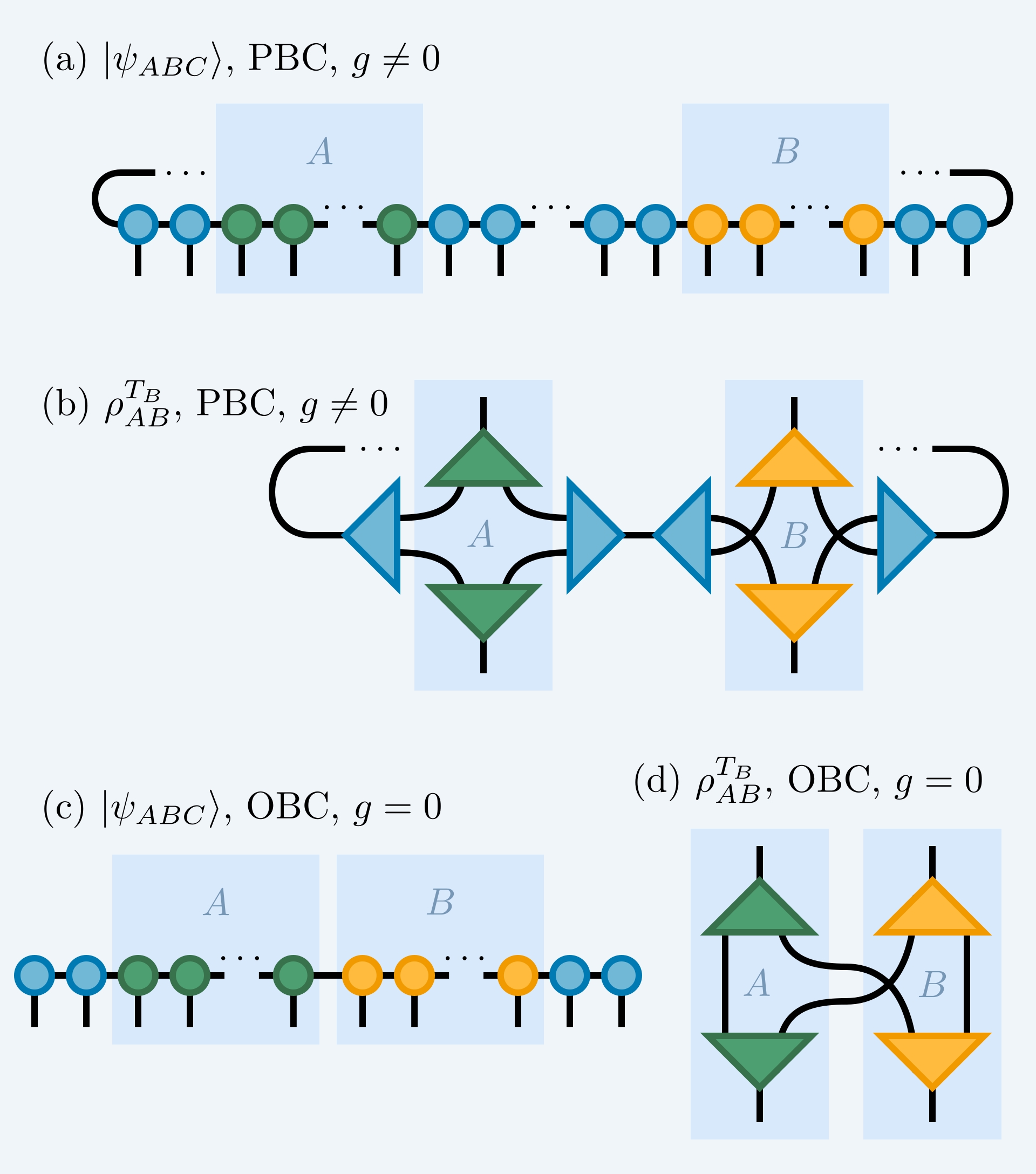}
    \caption{
    Forming a compressed representation of a partially transposed density matrix, \rhopt, derived from a MPS, $\ket{\psi_{ABC}}$, namely, lateral compression of all contiguous sections, followed by vertical decomposition of $A$ and $B$.
    (a) and (b): state and derived operator for PBC and non-zero gap, $g$, between contiguous subsystems $A$ and $B$ -- the most general geometry.
    (c) and (d): state and derived operator for OBC with zero gap between subsystems $A$ and $B$ -- a common geometry.
    The effective linear operator of \rhopt~ is formed by grouping the upper and lower indices of (b) and (d).
    }
    \label{fig:mps-compressed-rhopt}
\end{figure}

The above assumes nothing about the geometry of $A$ and $B$ within the MPS.
However, if we assume that $A$ and $B$ are contiguous blocks (as is often the case), then we can adapt the method for arbitrarily many sites by \emph{compressing} each block, a process sketched in Fig.~\ref{fig:mps-compress-section}.
First, a `lateral' compression of $A$, $B$, and, if necessary, any contiguous blocks of $C$ that form the environment.
This is the method derived in~\cite{pippan2010efficient} in order to efficiently address periodic boundary DMRG in the language of MPS.
Secondly, a `vertical' decomposition of subsystems $A$ and $B$ to reintroduce new effective physical indices to the density matrix~\cite{ruggiero2016entanglement,mbeng2017negativity}.
The details of the lateral compression as are follows:
\begin{enumerate}
  \item Form the transfer matrix of a contiguous section to be compressed -- Fig.~\ref{fig:mps-compress-section}(a)$\Rightarrow$(b).
  \item Perform a iterative SVD decomposition of the transfer matrix, treating it as a linear operator with effective dimensions $\chi^2 \times \chi^2$ by grouping the left and right bonds respectively -- Fig.~\ref{fig:mps-compress-section}(b)$\Rightarrow$(c).
  Note that generally, the longer a section is, the fewer the number of singular values required to represent its transfer matrix to high precision.
\end{enumerate}
The procedure for the `vertical' compression, which only is performed on subsystems $A$ and $B$ in order to reintroduce physical indices is as follows:
\begin{enumerate}
  \item Perform a decomposition of the section, which now might be in SVD form, but this time grouping the \emph{upper} and \emph{lower} bonds respectively -- Fig.~\ref{fig:mps-compress-section}(c)$\Rightarrow$(d).
  This operator, with effective dimensions $\chi^2 \times \chi^2$, is generally full-rank, however, we note that it is also positive symmetric, and thus the fast (compared to SVD) Cholesky decomposition can be used.
  \item `Split' the bond connecting the two symmetric factors simply by re-indexing the tensors -- Fig.~\ref{fig:mps-compress-section}(d)$\Rightarrow$(e).
  This re-introduces effective `ket' and `bra' physical indices to the section, with size $\leq \chi^2$ rather than exponential in the number of sites.
\end{enumerate}

With these two steps we have a method to derive a `compressed' representation of the partially traced, partially transposed density operator \rhopt, from a MPS with $A$ and $B$ contiguous blocks separated with gap $g$, as shown in Fig.~\ref{fig:mps-compressed-rhopt}(a).
First partition the state into $A$, $B$, and potentially several $C$ sections, then perform lateral compression on any of these that are long enough for it to make sense (e.g. $\chi^2 < p^{L_A}$ for section $A$).
Next, perform vertical decompositions on subsystems $A$ and $B$, and finally swap the `ket' and `bra' indices on subsystem $B$ to effect the partial transpose.
This resulting tensor network, in the most general geometry, is shown in Fig.~\ref{fig:mps-compressed-rhopt}(b).
The largest tensor it contains is always of size $\leq \chi^4$.

We note simplifications can be made to the network in several common scenarios.
If open boundary conditions (OBC) are used, the gauge freedom can be utilized to eliminate both the left and right environments completely.
Similarly, if either subsystem $A$ or $B$ contains the end of the chain, they can be represented as two identity tensors.
For periodic boundary conditions (PBC), the left and right environments are the \emph{same} section, and an effective gauge to eliminate them can only be introduced if the section's transfer matrix has a single dominant singular value -- i.e. it is separable.
Finally, clearly if $A$ and $B$ are adjacent ($g=0$), no environment is needed separate them.
Given this MPS-derived, compressed, tensor network representation of \rhopt, we can as before apply the SLQ method to this operator using a sample vector $\ket{\phi}$, also of size $\leq \chi^4$, to compute the logarithmic negativity of arbitrary contiguous sections.
As before, being able to contract the sample vector into the network to yield a new vector, rather than first contracting the full operator, yields the key efficiency saving.
The best contraction order depends on the various index dimensions, and in practice, we choose the order automatically using a greedy approach~\cite{smith2018opteinsum}.
Bond dimension now becomes the limiting factor of the algorithm, and if we translate the memory requirement of densely representing 30 qubits into this language, we find that $\chi \lesssim 180$ is the equivalent limit.

\section{Results \label{sec:results}}

We now move on to demonstrating the TNSLQ method in three different scenarios.
The first two results involve `partial trace states' -- random pure states and a many-body quench -- for which $L \leq 30$ and the entanglement varies from zero to highly-entangled.
The third studies the scaling of entanglement in a large matrix product state, namely, the ground-state of the Heisenberg Hamiltonian acquired using DMRG, for which analytic results are available.
All computations were performed using the open-source library \texttt{quimb}~\cite{gray2018quimb}, which has implementations of tensor network linear operators, the SLQ algorithm, and two-site DMRG.

\subsection{Random pure states}

\begin{figure}[tb]
  \centering
  \includegraphics[width=1\linewidth]{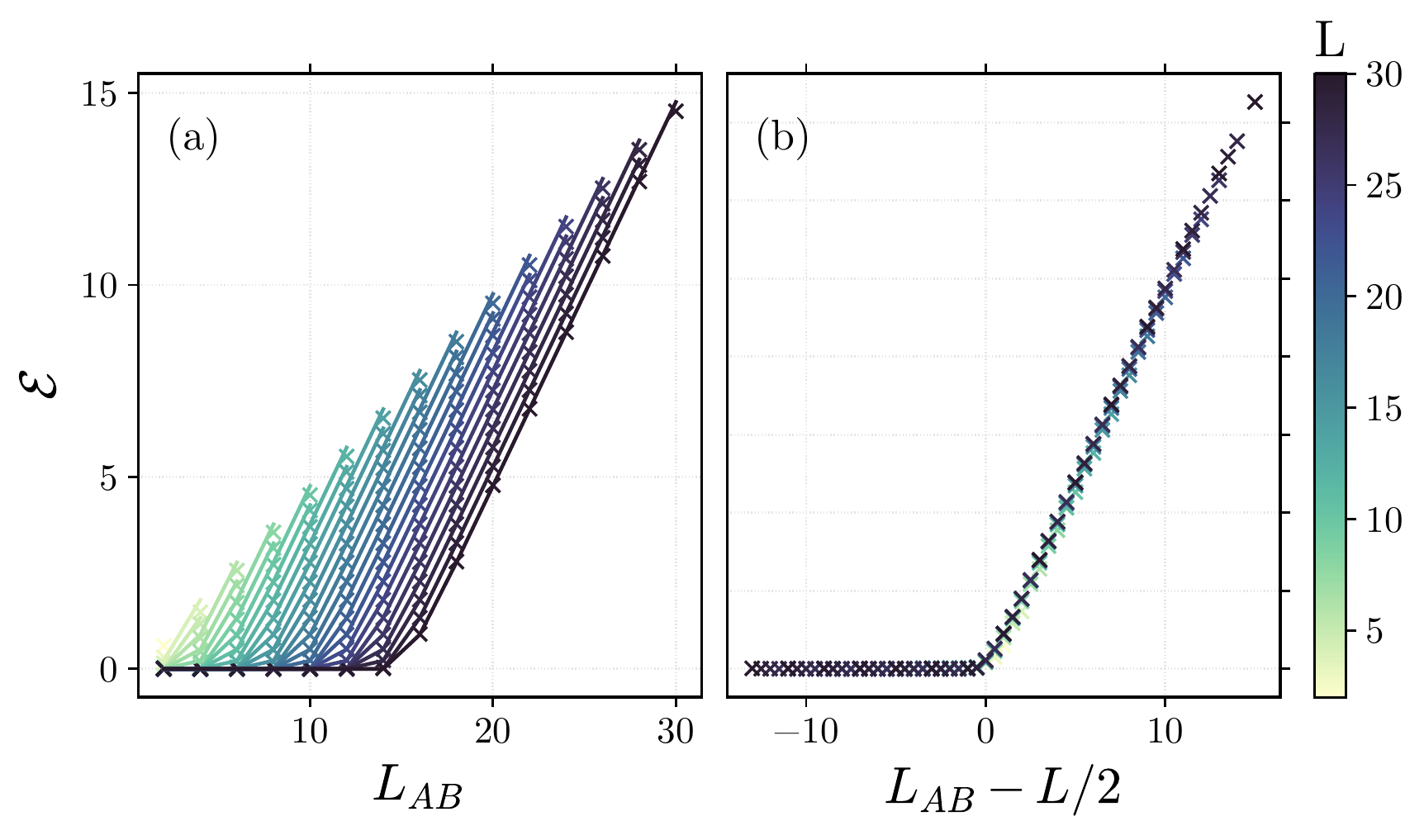}
  \caption{
  (a) Logarithmic negativity, $\mathcal{E}(\rho_{AB})$, between two equal subsystems of combined size $L_{AB}$ in a random pure state of total size $L$. The lines denote the analytical result, Eq.~\eqref{eq:ln-rand-analytic}, derived in~\cite{bhosale2012entanglement} while the crosses show results computed with TNSLQ. (b) The same but offset by $L/2$, where we see a clear collapse to universal behaviour across all lengths $L$.
  }
  \label{fig:random-logneg}
\end{figure}

First, we benchmark the TNSLQ method for density matrices derived from random pure states, in full dense representation, of length up to $L=30$ (total Hilbert space size $2^{30}=1073741824$).
Since these states are completely permutationally symmetric, the only variables are the size of $L_A$, $L_B$ and $L$.
We simply take $L_A=L_B=L_{AB} / 2$, then compute $\mathcal{E}$ for varying $L_{AB}$ and $L$.
For each configuration we average over 10 different random realizations, though there is very little variance between them.
The analytic result for these states is known~\cite{bhosale2012entanglement}:
\begin{equation} \label{eq:ln-rand-analytic}
  \mathcal{E}(\rho_{AB})
  =
  \log_2
  \left[
  \dfrac{2}{\pi} \sin^{-1} \left(\dfrac{1}{\tilde R} \right) +
  \dfrac{2 \left( 1 + 2 \tilde R ^ 2 \right) }{3 \pi \tilde R} \sqrt{1 - \dfrac{1}{\tilde R^2}} ~,
  \right]
\end{equation}
where $\tilde R = 2 \sqrt{d_A d_B / d_C}$, which we compare to in Fig.~\ref{fig:random-logneg}(a).
We find very good accordance with the analytic prediction, and confirm a universal behaviour whereby the entanglement is zero for $L_{AB}<L / 2$, and rises linear afterwards -- see Fig.~\ref{fig:random-logneg}(b).
We note that with the standard method of computing $\mathcal{E}$, approximately the right half of Fig.~\ref{fig:random-logneg}(a) ($L_{AB} \geq 15$) would not be available.

\subsection{Scrambling in a Quench}

\begin{figure}[tb]
  \centering
  \includegraphics[width=1\linewidth]{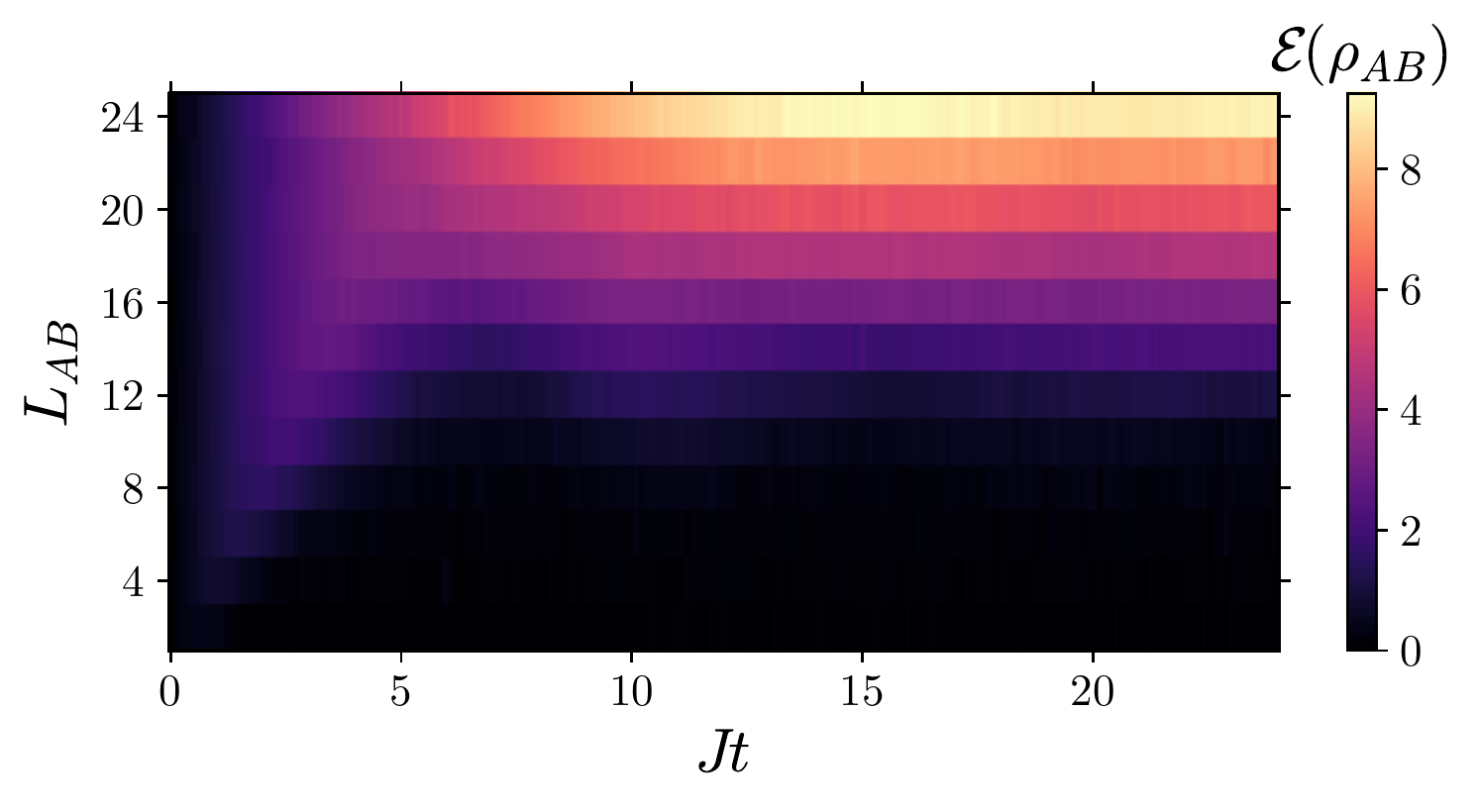}
  \caption{
  The entanglement, $\mathcal{E}(\rho_{AB})$, as computed with the TNSLQ method, across a central cut for a quench in the Heisenberg model as a function of time $Jt$ and reduced subsystem size $L_{AB}$
  The total chain length is $L=24$, and we take the reduced density operator $\rho_{AB}$ to be centered around the half way point such that either side there is environement of length $L - L_{AB} / 2$.
  }
  \label{fig:quench-L24}
\end{figure}

We next move on to applying the TNSLQ method to a more physical example - the time evolution of a state after a quench with an interacting many-body Hamiltonian.
We take a system of $L$ spin-1/2 particles with nearest neighbour Heisenberg Hamiltonian:
\begin{equation}\label{eq:heis-ham}
\hat{H}= J \sum_{i=1}^{L-1} \boldsymbol{\sigma}_i \cdot \boldsymbol{\sigma}_{i+1}~,
\end{equation}
where $J$ is the interaction strength and
$\boldsymbol{\sigma}_{i} = (\sigma^x_i, \sigma^y_i, \sigma^z_i)$
the vector of spin operators matrices acting on site $i$.
The system is initialized in the (separable)
Neel-state $\ket{\Psi(0)} = \ket{\uparrow \downarrow \uparrow \dots}$ and evolved using integration according to the equation
$\frac{\partial}{\partial_t} \ket{\psi(t)} = - i \hat{H} \ket{\psi(t)}$, where we have set the Planck constant $\hbar = 1$.
In terms of geometry, we choose $A$ and $B$ as neighbouring blocks of equal length, either side of a central cut in the chain, of total length $L_{AB}$.
In Fig.~\ref{fig:quench-L24} we plot the logarithmic negativity, $\mathcal{E}(\rho_{AB})$, computed using TNSLQ, for this set-up as a function of time $t$, for a chain of total length $L=24$.
Again, approximately the upper half of this figure would not be computable using the exact method of calculating $\mathcal{E}$, but for those sizes that are, we find very good accordance (not shown) within the target precision of 1\% for the TNSLQ method.
In relation to scrambling~\cite{sekino2008fast}, we expect information describing the initial system to quickly de-localize, building up entanglement at increasingly longer scales.
This also means that for sufficiently short length-scales, the entanglement should grow and \emph{then} decrease, as the combined subsystem $\rho_{AB}$ becomes increasingly entangled with $C$, precluding entanglement between $A$ and $B$.
This is exactly what we see in Fig.~\ref{fig:quench-L24}, entanglement growing and then dying at increasing length-scales, such that a subsystem of size $L_{AB}$ is `scrambled' after time $L_{AB} / 2$.
Eventually the system `equilibrates' with an entanglement structure similar to that of a random state -- zero entanglement if $L_{AB} < L / 2$, then rising roughly linearly as $L_{AB}\rightarrow L$.

\subsection{Heisenberg ground-state}

\begin{figure}[tb]
  \centering
  \includegraphics[width=\linewidth]{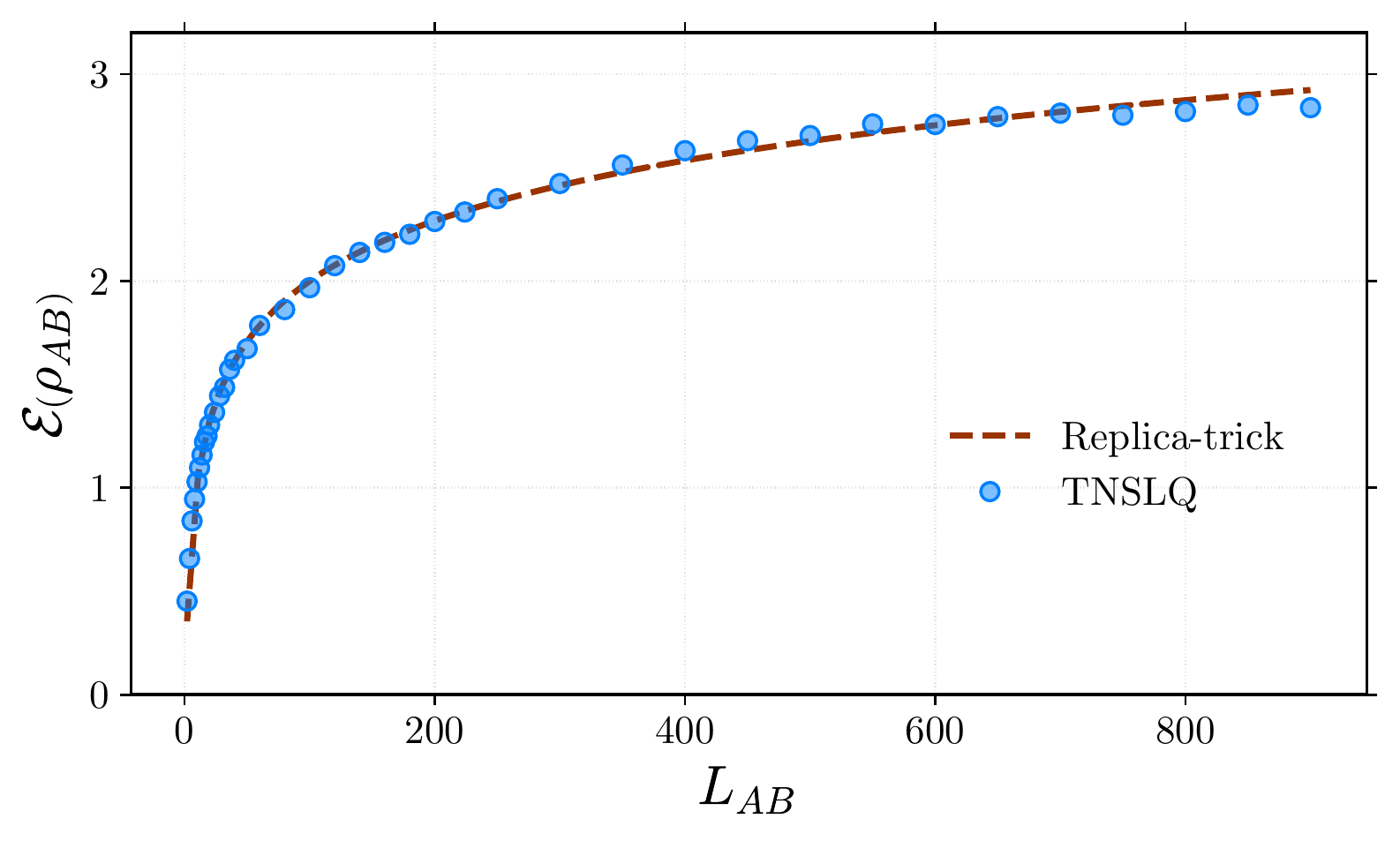}
  \caption{
  Logarithmic negativity, $\mathcal{E}$, between two neighbouring contiguous blocks of total size, $L_{AB}$, at the center of system of total length $L=1000$.
  Blue markers show the entanglement computed using the TNSLQ method.
  The red dashed line shows a fit (using data where $L_{AB} \leq L / 2$) to the conformal field theory prediction as given in Eq.~\eqref{eq:heis-CFT-predic}.
  }
  \label{fig:heis-logneg}
\end{figure}

Finally, we demonstrate the TNSLQ method for MPS subsystems by studying the logarithmic negativity for two adjacent blocks in the ground-state of the Heisenberg model, with Hamiltonian as defined in Eq.~\eqref{eq:heis-ham}.
The analytic form of this has been derived using conformal field theory and the `replica trick'~\cite{calabrese2009entanglement,calabrese2012entanglement,calabrese2013entanglement}.
The behaviour, which is universal, is logarithmic scaling of the entanglement with block size, as given by
\begin{equation}\label{eq:heis-CFT-predic}
  \mathcal{E}(\rho_{AB}) = \frac{c}{4} \log_2 \left( \frac{L_{AB}}{4} \right) + K
\end{equation}
for central charge $c$ and constant $K$.
We take $L=1000$ then generate a MPS description of the ground-state of $\hat{H}$ using two-site DMRG~\cite{white1992density,schollwock2011density}.
The state has a maximum bond dimension of $\chi=116$ at the center, with $\chi$ also remaining above 90 for $\sim 80\%$ of the chain.
Since we are using OBC and neighbouring blocks, so that $g=0$, the form of the compressed version of \rhopt~ is that of Fig.~\ref{fig:mps-compressed-rhopt}(d).
The TNSLQ computed entanglement, $\mathcal{E}(\rho_{AB})$, is shown as a function of $L_{AB}$ in Fig.~\ref{fig:heis-logneg}.
Also plotted is a fit of Eq.~\eqref{eq:heis-CFT-predic} -- found using $L_{AB} \leq 500$ to avoid finite size effects.
The results are very closely matching, with possibly a slight trend below logarithmic growth for long $L_{AB}$, which we attribute to $L_{AB}$ approaching $L$.
From the fit of Eq.~\eqref{eq:heis-CFT-predic} we find
$c = 1.16595
\pm  0.00001$ and
$K = 0.6458
\pm  0.0001
$

\section{Error Analysis} \label{sec:error-analysis}

The TNSLQ method is fundamentally a stochastic process and thus comes with a certain limitation on achieving very high precision estimates.
In fact, the effort scales exponentially with the number of decimal places required~\cite{avron2011randomized}.
Crucially however, and as shown by our results above, a constant level of precision of 0.1 - 1\% is easily achievable, and for many simulation purposes, completely sufficient.
Moreover, the error on the estimate is easy to keep track of.
To put this on more concrete terms, consider that the variance of a single estimate, $G_n$, of operator $\hat Y = f(\hat X)$, using Hutchinson's trace method is bounded by~\cite{hutchinson1990stochastic,avron2011randomized}:
\begin{equation} \label{eq:hutchinson-single-variance}
  \text{Var}(G_n) = 2 \tr (\hat Y^\dag \hat Y) - 2 \sum_i \hat Y^2_{ii}~.
\end{equation}
Ignoring the second term, which is strictly negative and thus beneficial, we can assess the first for $\hat Y \rightarrow |\rhopt|$.
Since \rhopt~ is Hermitian it follows that:
\begin{align} \label{eq:rhopt-slq-var}
  \tr (\hat Y^\dag \hat Y) &= \tr (|\rhopt| |\rhopt|) \nonumber\\
                           &= \tr ((\rhopt)^2) \nonumber \\
                           &= \tr (\rho_{AB}^2)
\end{align}
which is simply the \emph{purity} of the joint density matrix, whose value lies between $(d_A d_B)^{-1}$ and 1.
Interestingly, this implies that the entanglement will be easier to compute the \emph{more} mixed a state $\rho_{AB}$ is.
Even more importantly, the upper limit on the variance is constant.
By substituting Eq.~\eqref{eq:rhopt-slq-var} into Eq.~\eqref{eq:hutchinson-single-variance} we find $\text{Var}(G_n) \leq 2$, and thus there are no hidden costs of scaling to larger system sizes.
Additionally, as one estimates a quantity with TNSLQ, the error on the estimate can be tracked simply as the standard error on the mean of the $N$ actual estimates computed so far, $\{G_1, G_2, \ldots, G_N\}$, yielding
estimated error $\sqrt{\text{Var}(\{G_i\}) / N }$.
As such, the TNSLQ method for quantities based on $\rho_{AB}$ yield errors which have the desirable properties of being both well-controlled and readily accessible.

\section{Discussion} \label{sec:discussion}

We have seen that the TNSLQ method enables the fast computation of many-body entanglement for various states.
It involves treating a tensor network representation of the partially traced, partially transposed density matrix, \rhopt\, as a linear operator.
The action of this operator can be efficiently evaluated by contracting the sample vector into it, allowing one to use the SLQ procedure to compute any quantity of the form $\tr{f(A)}$.
The entanglement negativity is one such quantity when we set $A=\rhopt$ and $f=\text{abs}$.
We note that since generally these operators do not have equivalent efficient representations as either sparse matrices or low-rank operators, both the tensor network description and SLQ procedure seem necessary components.
We have focussed particularly on the entanglement of bipartite density matrices derived as subsystems of larger, pure states, but we note that the method should be just as applicable to the situation where one begins with an efficient tensor network representation of a mixed state.

For pure states represented as vectors in their full Hilbert space, the TNSLQ method enables the efficient computation of the logarithmic negativity between subsystems $A$ and $B$ for \emph{any} tri-partition of $\ket{\psi}$, that is, with any choice of environment $C$.
Roughly speaking, in many simulations this doubles the size $L_{AB}$ for which it is tractable to compute $\mathcal{E}(\rho_{AB})$.
For MPS, we combined a method of efficiently finding a compressed form of the bipartite reduced density matrix, $\rho_{AB}$, with the SLQ procedure used to then find the logarithmic negativity.
The method is tractable for OBC or PBC as well as disjoint blocks separated by length $g$.
We note that unlike some previous studies, at no point do we have to arbitrarily curtail the number of states kept (which likely introduces a systematic error), as long as the initial bond size, $\chi \lesssim 180$.
To move beyond this limit, it might be worth exploring the actual effect of limiting certain bond sizes, for example in the vertical decomposition of the $A$ and $B$ subsystem sections into symmetric Cholesky factors.
Another interesting avenue is whether one could store the Lanczos sampling vectors $\{\ket{\phi_n}\}$ in an efficient form - an obvious choice being as MPS.
In this case, during the SLQ procedure the bond dimension would steadily rise, probably requiring the restriction to a fixed bond size manifold -- how this might bias the estimate is not clear.
A relevant approach was taken recently in~\cite{august2017approximation}, where both the target operator $X$ is a \emph{matrix product operator} (MPO), as well as sampling unitaries used to perform a block Lanczos procedure.
While it is simple to form a MPO representation of $\rhopt$ from a MPS, this has an increased storage cost $L p^2 \chi^4$, and the bond dimension of the sampling unitaries must also be artificially restricted, making the MPO approach potentially unsuitable in this particular instance.

The TNSLQ method is easily capable of estimating quantities to the level of 0.1-1\%, but, as a fundamentally stochastic process, it might not be suitable for computing quantities to many digits of precision.
On the other hand, we have shown that the error for the density operator $|\rhopt|$, in terms of the variance of individual estimates, is both well controlled - being bounded by a system size independent constant - and easy to keep track of.
Moreover, this feature of being an average over many low-precision estimates, as well as having a low memory-overhead, makes the TNSLQ method easy to accelerate.
Firstly, it is trivial to parallelize the algorithm over independent random estimates.
Secondly, single precision arithmetic can be used, for which graphical processing units (GPUs) are particularly suited.
Implementations of both of these accelerations have been incorporated into the open-source library \texttt{quimb}~\cite{gray2018quimb}.

Finally, we note that although the logarithmic negativity of `partial trace states' and matrix product states are particular instances of TNSLQ, there are plenty of other potential candidates for the both the tensor network operator and computed function $\tr{f(\cdot)}$.
For instance, the Von Neumann entropy is given by $-\tr{\rho \log_2 \rho}$.
It is easy to simplify the PTS and MPS procedures presented above, by removing the partial transpose and considering a single subsystem only, to compute this and hence, for example, the mutual information.
Other suitable functions include the partition function, $Z=\tr(e^{-\beta \hat{H}})$ for tensor network Hamiltonian $\hat{H}$, and the Frobenius norm $||\hat X||_F = \sqrt{\tr(\hat X^2)}$ for Hermitian tensor network $\hat X$.
In terms of other many-body quantum states, the TNSLQ method should be trivially applicable to tree tensor networks~\cite{shi2006classical,murg2010simulating} and multi-scale entanglement renormalization ansatz states~\cite{vidal2007entanglement,vidal2008class}.
Furthermore, state of the art classical simulations of quantum computation have also recently relied on tensor network descriptions of the full circuit\cite{markov2008simulating,pednault2017breaking,boixo2017simulation,chen2018classical,markov2018quantum}.
Even without developing any compression schemes specific to these structures, the computation of $\mathcal{E}(\rho_{AB})$ for $L_{AB}$ up to $\sim30$ should now be possible.
Needless to say, in all the above cases there are many interesting questions that might be probed with a genuine, many-body entanglement measure such as the logarithmic negativity.

\emph{Acknowledgements.-- }
JG acknowledges funding from the EPSRC Center for Doctoral Training in Delivering Quantum Technologies at UCL.

\bibliography{recln-main}

\begin{thebibliography}{60}%
\makeatletter
\providecommand \@ifxundefined [1]{%
 \@ifx{#1\undefined}
}%
\providecommand \@ifnum [1]{%
 \ifnum #1\expandafter \@firstoftwo
 \else \expandafter \@secondoftwo
 \fi
}%
\providecommand \@ifx [1]{%
 \ifx #1\expandafter \@firstoftwo
 \else \expandafter \@secondoftwo
 \fi
}%
\providecommand \natexlab [1]{#1}%
\providecommand \enquote  [1]{``#1''}%
\providecommand \bibnamefont  [1]{#1}%
\providecommand \bibfnamefont [1]{#1}%
\providecommand \citenamefont [1]{#1}%
\providecommand \href@noop [0]{\@secondoftwo}%
\providecommand \href [0]{\begingroup \@sanitize@url \@href}%
\providecommand \@href[1]{\@@startlink{#1}\@@href}%
\providecommand \@@href[1]{\endgroup#1\@@endlink}%
\providecommand \@sanitize@url [0]{\catcode `\\12\catcode `\$12\catcode
  `\&12\catcode `\#12\catcode `\^12\catcode `\_12\catcode `\%12\relax}%
\providecommand \@@startlink[1]{}%
\providecommand \@@endlink[0]{}%
\providecommand \url  [0]{\begingroup\@sanitize@url \@url }%
\providecommand \@url [1]{\endgroup\@href {#1}{\urlprefix }}%
\providecommand \urlprefix  [0]{URL }%
\providecommand \Eprint [0]{\href }%
\providecommand \doibase [0]{http://dx.doi.org/}%
\providecommand \selectlanguage [0]{\@gobble}%
\providecommand \bibinfo  [0]{\@secondoftwo}%
\providecommand \bibfield  [0]{\@secondoftwo}%
\providecommand \translation [1]{[#1]}%
\providecommand \BibitemOpen [0]{}%
\providecommand \bibitemStop [0]{}%
\providecommand \bibitemNoStop [0]{.\EOS\space}%
\providecommand \EOS [0]{\spacefactor3000\relax}%
\providecommand \BibitemShut  [1]{\csname bibitem#1\endcsname}%
\let\auto@bib@innerbib\@empty
\bibitem [{\citenamefont {Rao}(1945)}]{rao1945information}%
  \BibitemOpen
  \bibfield  {author} {\bibinfo {author} {\bibfnamefont {C.~R.}\ \bibnamefont
  {Rao}},\ }\href@noop {} {\bibfield  {journal} {\bibinfo  {journal} {Bull.
  Calcutta Math. Soc}\ }\textbf {\bibinfo {volume} {37}},\ \bibinfo {pages}
  {81} (\bibinfo {year} {1945})}\BibitemShut {NoStop}%
\bibitem [{\citenamefont {Shor}(1999)}]{shor1999polynomial}%
  \BibitemOpen
  \bibfield  {author} {\bibinfo {author} {\bibfnamefont {P.~W.}\ \bibnamefont
  {Shor}},\ }\href@noop {} {\bibfield  {journal} {\bibinfo  {journal} {SIAM
  Rev.}\ }\textbf {\bibinfo {volume} {41}},\ \bibinfo {pages} {303} (\bibinfo
  {year} {1999})}\BibitemShut {NoStop}%
\bibitem [{\citenamefont {Harrow}\ \emph {et~al.}(2009)\citenamefont {Harrow},
  \citenamefont {Hassidim},\ and\ \citenamefont {Lloyd}}]{harrow2009quantum}%
  \BibitemOpen
  \bibfield  {author} {\bibinfo {author} {\bibfnamefont {A.~W.}\ \bibnamefont
  {Harrow}}, \bibinfo {author} {\bibfnamefont {A.}~\bibnamefont {Hassidim}}, \
  and\ \bibinfo {author} {\bibfnamefont {S.}~\bibnamefont {Lloyd}},\
  }\href@noop {} {\bibfield  {journal} {\bibinfo  {journal} {Phys. Rev. Lett.}\
  }\textbf {\bibinfo {volume} {103}},\ \bibinfo {pages} {150502} (\bibinfo
  {year} {2009})}\BibitemShut {NoStop}%
\bibitem [{\citenamefont {Bennett}\ \emph {et~al.}(1993)\citenamefont
  {Bennett}, \citenamefont {Brassard}, \citenamefont {Cr{\'e}peau},
  \citenamefont {Jozsa}, \citenamefont {Peres},\ and\ \citenamefont
  {Wootters}}]{bennett1993teleporting}%
  \BibitemOpen
  \bibfield  {author} {\bibinfo {author} {\bibfnamefont {C.~H.}\ \bibnamefont
  {Bennett}}, \bibinfo {author} {\bibfnamefont {G.}~\bibnamefont {Brassard}},
  \bibinfo {author} {\bibfnamefont {C.}~\bibnamefont {Cr{\'e}peau}}, \bibinfo
  {author} {\bibfnamefont {R.}~\bibnamefont {Jozsa}}, \bibinfo {author}
  {\bibfnamefont {A.}~\bibnamefont {Peres}}, \ and\ \bibinfo {author}
  {\bibfnamefont {W.~K.}\ \bibnamefont {Wootters}},\ }\href@noop {} {\bibfield
  {journal} {\bibinfo  {journal} {Phys. Rev. Lett.}\ }\textbf {\bibinfo
  {volume} {70}},\ \bibinfo {pages} {1895} (\bibinfo {year}
  {1993})}\BibitemShut {NoStop}%
\bibitem [{\citenamefont {Bennett}\ and\ \citenamefont
  {Wiesner}(1992)}]{bennett1992communication}%
  \BibitemOpen
  \bibfield  {author} {\bibinfo {author} {\bibfnamefont {C.~H.}\ \bibnamefont
  {Bennett}}\ and\ \bibinfo {author} {\bibfnamefont {S.~J.}\ \bibnamefont
  {Wiesner}},\ }\href@noop {} {\bibfield  {journal} {\bibinfo  {journal} {Phys.
  Rev. Lett.}\ }\textbf {\bibinfo {volume} {69}},\ \bibinfo {pages} {2881}
  (\bibinfo {year} {1992})}\BibitemShut {NoStop}%
\bibitem [{\citenamefont {Ekert}(1991)}]{ekert1991quantum}%
  \BibitemOpen
  \bibfield  {author} {\bibinfo {author} {\bibfnamefont {A.~K.}\ \bibnamefont
  {Ekert}},\ }\href@noop {} {\bibfield  {journal} {\bibinfo  {journal} {Phys.
  Rev. Lett.}\ }\textbf {\bibinfo {volume} {67}},\ \bibinfo {pages} {661}
  (\bibinfo {year} {1991})}\BibitemShut {NoStop}%
\bibitem [{\citenamefont {Amico}\ \emph {et~al.}(2008)\citenamefont {Amico},
  \citenamefont {Fazio}, \citenamefont {Osterloh},\ and\ \citenamefont
  {Vedral}}]{amico2008entanglement}%
  \BibitemOpen
  \bibfield  {author} {\bibinfo {author} {\bibfnamefont {L.}~\bibnamefont
  {Amico}}, \bibinfo {author} {\bibfnamefont {R.}~\bibnamefont {Fazio}},
  \bibinfo {author} {\bibfnamefont {A.}~\bibnamefont {Osterloh}}, \ and\
  \bibinfo {author} {\bibfnamefont {V.}~\bibnamefont {Vedral}},\ }\href@noop {}
  {\bibfield  {journal} {\bibinfo  {journal} {Rev. Mod. Phys.}\ }\textbf
  {\bibinfo {volume} {80}},\ \bibinfo {pages} {517} (\bibinfo {year}
  {2008})}\BibitemShut {NoStop}%
\bibitem [{\citenamefont {Schollw{\"o}ck}(2011)}]{schollwock2011density}%
  \BibitemOpen
  \bibfield  {author} {\bibinfo {author} {\bibfnamefont {U.}~\bibnamefont
  {Schollw{\"o}ck}},\ }\href@noop {} {\bibfield  {journal} {\bibinfo  {journal}
  {Ann. Phys.}\ }\textbf {\bibinfo {volume} {326}},\ \bibinfo {pages} {96}
  (\bibinfo {year} {2011})}\BibitemShut {NoStop}%
\bibitem [{\citenamefont {Calabrese}\ \emph {et~al.}(2012)\citenamefont
  {Calabrese}, \citenamefont {Cardy},\ and\ \citenamefont
  {Tonni}}]{calabrese2012entanglement}%
  \BibitemOpen
  \bibfield  {author} {\bibinfo {author} {\bibfnamefont {P.}~\bibnamefont
  {Calabrese}}, \bibinfo {author} {\bibfnamefont {J.}~\bibnamefont {Cardy}}, \
  and\ \bibinfo {author} {\bibfnamefont {E.}~\bibnamefont {Tonni}},\
  }\href@noop {} {\bibfield  {journal} {\bibinfo  {journal} {Phys. Rev. Lett.}\
  }\textbf {\bibinfo {volume} {109}},\ \bibinfo {pages} {130502} (\bibinfo
  {year} {2012})}\BibitemShut {NoStop}%
\bibitem [{\citenamefont {Nishioka}\ \emph {et~al.}(2009)\citenamefont
  {Nishioka}, \citenamefont {Ryu},\ and\ \citenamefont
  {Takayanagi}}]{nishioka2009holographic}%
  \BibitemOpen
  \bibfield  {author} {\bibinfo {author} {\bibfnamefont {T.}~\bibnamefont
  {Nishioka}}, \bibinfo {author} {\bibfnamefont {S.}~\bibnamefont {Ryu}}, \
  and\ \bibinfo {author} {\bibfnamefont {T.}~\bibnamefont {Takayanagi}},\
  }\href@noop {} {\bibfield  {journal} {\bibinfo  {journal} {Journal of Physics
  A: Mathematical and Theoretical}\ }\textbf {\bibinfo {volume} {42}},\
  \bibinfo {pages} {504008} (\bibinfo {year} {2009})}\BibitemShut {NoStop}%
\bibitem [{\citenamefont {Calabrese}\ and\ \citenamefont
  {Cardy}(2009)}]{calabrese2009entanglement}%
  \BibitemOpen
  \bibfield  {author} {\bibinfo {author} {\bibfnamefont {P.}~\bibnamefont
  {Calabrese}}\ and\ \bibinfo {author} {\bibfnamefont {J.}~\bibnamefont
  {Cardy}},\ }\href@noop {} {\bibfield  {journal} {\bibinfo  {journal} {J.
  Phys. A}\ }\textbf {\bibinfo {volume} {42}},\ \bibinfo {pages} {504005}
  (\bibinfo {year} {2009})}\BibitemShut {NoStop}%
\bibitem [{\citenamefont {Wichterich}\ \emph {et~al.}(2009)\citenamefont
  {Wichterich}, \citenamefont {Molina-Vilaplana},\ and\ \citenamefont
  {Bose}}]{wichterich2009scaling}%
  \BibitemOpen
  \bibfield  {author} {\bibinfo {author} {\bibfnamefont {H.}~\bibnamefont
  {Wichterich}}, \bibinfo {author} {\bibfnamefont {J.}~\bibnamefont
  {Molina-Vilaplana}}, \ and\ \bibinfo {author} {\bibfnamefont
  {S.}~\bibnamefont {Bose}},\ }\href@noop {} {\bibfield  {journal} {\bibinfo
  {journal} {Physical Review A}\ }\textbf {\bibinfo {volume} {80}},\ \bibinfo
  {pages} {010304} (\bibinfo {year} {2009})}\BibitemShut {NoStop}%
\bibitem [{\citenamefont {Bayat}\ \emph {et~al.}(2010)\citenamefont {Bayat},
  \citenamefont {Sodano},\ and\ \citenamefont {Bose}}]{bayat2010negativity}%
  \BibitemOpen
  \bibfield  {author} {\bibinfo {author} {\bibfnamefont {A.}~\bibnamefont
  {Bayat}}, \bibinfo {author} {\bibfnamefont {P.}~\bibnamefont {Sodano}}, \
  and\ \bibinfo {author} {\bibfnamefont {S.}~\bibnamefont {Bose}},\ }\href@noop
  {} {\bibfield  {journal} {\bibinfo  {journal} {Phys. Rev. B}\ }\textbf
  {\bibinfo {volume} {81}},\ \bibinfo {pages} {064429} (\bibinfo {year}
  {2010})}\BibitemShut {NoStop}%
\bibitem [{\citenamefont {Caruso}\ \emph {et~al.}(2010)\citenamefont {Caruso},
  \citenamefont {Chin}, \citenamefont {Datta}, \citenamefont {Huelga},\ and\
  \citenamefont {Plenio}}]{caruso2010entanglement}%
  \BibitemOpen
  \bibfield  {author} {\bibinfo {author} {\bibfnamefont {F.}~\bibnamefont
  {Caruso}}, \bibinfo {author} {\bibfnamefont {A.~W.}\ \bibnamefont {Chin}},
  \bibinfo {author} {\bibfnamefont {A.}~\bibnamefont {Datta}}, \bibinfo
  {author} {\bibfnamefont {S.~F.}\ \bibnamefont {Huelga}}, \ and\ \bibinfo
  {author} {\bibfnamefont {M.~B.}\ \bibnamefont {Plenio}},\ }\href@noop {}
  {\bibfield  {journal} {\bibinfo  {journal} {Phys. Rev. A}\ }\textbf {\bibinfo
  {volume} {81}},\ \bibinfo {pages} {062346} (\bibinfo {year}
  {2010})}\BibitemShut {NoStop}%
\bibitem [{\citenamefont {Calabrese}\ \emph {et~al.}(2013)\citenamefont
  {Calabrese}, \citenamefont {Cardy},\ and\ \citenamefont
  {Tonni}}]{calabrese2013entanglement}%
  \BibitemOpen
  \bibfield  {author} {\bibinfo {author} {\bibfnamefont {P.}~\bibnamefont
  {Calabrese}}, \bibinfo {author} {\bibfnamefont {J.}~\bibnamefont {Cardy}}, \
  and\ \bibinfo {author} {\bibfnamefont {E.}~\bibnamefont {Tonni}},\
  }\href@noop {} {\bibfield  {journal} {\bibinfo  {journal} {Journal of
  Statistical Mechanics: Theory and Experiment}\ }\textbf {\bibinfo {volume}
  {2013}},\ \bibinfo {pages} {P02008} (\bibinfo {year} {2013})}\BibitemShut
  {NoStop}%
\bibitem [{\citenamefont {Eisler}\ and\ \citenamefont
  {Zimbor{\'a}s}(2014)}]{eisler2014entanglement}%
  \BibitemOpen
  \bibfield  {author} {\bibinfo {author} {\bibfnamefont {V.}~\bibnamefont
  {Eisler}}\ and\ \bibinfo {author} {\bibfnamefont {Z.}~\bibnamefont
  {Zimbor{\'a}s}},\ }\href@noop {} {\bibfield  {journal} {\bibinfo  {journal}
  {New J. Phys.}\ }\textbf {\bibinfo {volume} {16}},\ \bibinfo {pages} {123020}
  (\bibinfo {year} {2014})}\BibitemShut {NoStop}%
\bibitem [{\citenamefont {Wen}\ \emph {et~al.}(2015)\citenamefont {Wen},
  \citenamefont {Chang},\ and\ \citenamefont {Ryu}}]{wen2015entanglement}%
  \BibitemOpen
  \bibfield  {author} {\bibinfo {author} {\bibfnamefont {X.}~\bibnamefont
  {Wen}}, \bibinfo {author} {\bibfnamefont {P.-Y.}\ \bibnamefont {Chang}}, \
  and\ \bibinfo {author} {\bibfnamefont {S.}~\bibnamefont {Ryu}},\ }\href@noop
  {} {\bibfield  {journal} {\bibinfo  {journal} {Phys. Rev. B}\ }\textbf
  {\bibinfo {volume} {92}},\ \bibinfo {pages} {075109} (\bibinfo {year}
  {2015})}\BibitemShut {NoStop}%
\bibitem [{\citenamefont {Sherman}\ \emph {et~al.}(2016)\citenamefont
  {Sherman}, \citenamefont {Devakul}, \citenamefont {Hastings},\ and\
  \citenamefont {Singh}}]{sherman2016nonzero}%
  \BibitemOpen
  \bibfield  {author} {\bibinfo {author} {\bibfnamefont {N.~E.}\ \bibnamefont
  {Sherman}}, \bibinfo {author} {\bibfnamefont {T.}~\bibnamefont {Devakul}},
  \bibinfo {author} {\bibfnamefont {M.~B.}\ \bibnamefont {Hastings}}, \ and\
  \bibinfo {author} {\bibfnamefont {R.~R.}\ \bibnamefont {Singh}},\ }\href@noop
  {} {\bibfield  {journal} {\bibinfo  {journal} {Physical Review E}\ }\textbf
  {\bibinfo {volume} {93}},\ \bibinfo {pages} {022128} (\bibinfo {year}
  {2016})}\BibitemShut {NoStop}%
\bibitem [{\citenamefont {Bayat}(2017)}]{bayat2017scaling}%
  \BibitemOpen
  \bibfield  {author} {\bibinfo {author} {\bibfnamefont {A.}~\bibnamefont
  {Bayat}},\ }\href@noop {} {\bibfield  {journal} {\bibinfo  {journal}
  {Physical review letters}\ }\textbf {\bibinfo {volume} {118}},\ \bibinfo
  {pages} {036102} (\bibinfo {year} {2017})}\BibitemShut {NoStop}%
\bibitem [{\citenamefont {Gray}\ \emph {et~al.}(2018)\citenamefont {Gray},
  \citenamefont {Bose},\ and\ \citenamefont {Bayat}}]{gray2018many}%
  \BibitemOpen
  \bibfield  {author} {\bibinfo {author} {\bibfnamefont {J.}~\bibnamefont
  {Gray}}, \bibinfo {author} {\bibfnamefont {S.}~\bibnamefont {Bose}}, \ and\
  \bibinfo {author} {\bibfnamefont {A.}~\bibnamefont {Bayat}},\ }\href@noop {}
  {\bibfield  {journal} {\bibinfo  {journal} {Phys. Rev. B}\ }\textbf {\bibinfo
  {volume} {97}},\ \bibinfo {pages} {201105} (\bibinfo {year}
  {2018})}\BibitemShut {NoStop}%
\bibitem [{\citenamefont {Huang}(2014)}]{huang2014computing}%
  \BibitemOpen
  \bibfield  {author} {\bibinfo {author} {\bibfnamefont {Y.}~\bibnamefont
  {Huang}},\ }\href@noop {} {\bibfield  {journal} {\bibinfo  {journal} {New
  journal of physics}\ }\textbf {\bibinfo {volume} {16}},\ \bibinfo {pages}
  {033027} (\bibinfo {year} {2014})}\BibitemShut {NoStop}%
\bibitem [{\citenamefont {{\.Z}yczkowski}\ \emph {et~al.}(1998)\citenamefont
  {{\.Z}yczkowski}, \citenamefont {Horodecki}, \citenamefont {Sanpera},\ and\
  \citenamefont {Lewenstein}}]{zyczkowski1998volume}%
  \BibitemOpen
  \bibfield  {author} {\bibinfo {author} {\bibfnamefont {K.}~\bibnamefont
  {{\.Z}yczkowski}}, \bibinfo {author} {\bibfnamefont {P.}~\bibnamefont
  {Horodecki}}, \bibinfo {author} {\bibfnamefont {A.}~\bibnamefont {Sanpera}},
  \ and\ \bibinfo {author} {\bibfnamefont {M.}~\bibnamefont {Lewenstein}},\
  }\href@noop {} {\bibfield  {journal} {\bibinfo  {journal} {Phys. Rev. A}\
  }\textbf {\bibinfo {volume} {58}},\ \bibinfo {pages} {883} (\bibinfo {year}
  {1998})}\BibitemShut {NoStop}%
\bibitem [{\citenamefont {Lee}\ \emph {et~al.}(2000)\citenamefont {Lee},
  \citenamefont {Kim}, \citenamefont {Park},\ and\ \citenamefont
  {Lee}}]{lee2000partial}%
  \BibitemOpen
  \bibfield  {author} {\bibinfo {author} {\bibfnamefont {J.}~\bibnamefont
  {Lee}}, \bibinfo {author} {\bibfnamefont {M.}~\bibnamefont {Kim}}, \bibinfo
  {author} {\bibfnamefont {Y.}~\bibnamefont {Park}}, \ and\ \bibinfo {author}
  {\bibfnamefont {S.}~\bibnamefont {Lee}},\ }\href@noop {} {\bibfield
  {journal} {\bibinfo  {journal} {J. Mod. Opt.}\ }\textbf {\bibinfo {volume}
  {47}},\ \bibinfo {pages} {2151} (\bibinfo {year} {2000})}\BibitemShut
  {NoStop}%
\bibitem [{\citenamefont {Vidal}\ and\ \citenamefont
  {Werner}(2002)}]{vidal2002computable}%
  \BibitemOpen
  \bibfield  {author} {\bibinfo {author} {\bibfnamefont {G.}~\bibnamefont
  {Vidal}}\ and\ \bibinfo {author} {\bibfnamefont {R.~F.}\ \bibnamefont
  {Werner}},\ }\href@noop {} {\bibfield  {journal} {\bibinfo  {journal} {Phys.
  Rev. A}\ }\textbf {\bibinfo {volume} {65}},\ \bibinfo {pages} {032314}
  (\bibinfo {year} {2002})}\BibitemShut {NoStop}%
\bibitem [{\citenamefont {Plenio}(2005)}]{plenio2005logarithmic}%
  \BibitemOpen
  \bibfield  {author} {\bibinfo {author} {\bibfnamefont {M.~B.}\ \bibnamefont
  {Plenio}},\ }\href@noop {} {\bibfield  {journal} {\bibinfo  {journal} {Phys.
  Rev. Lett.}\ }\textbf {\bibinfo {volume} {95}},\ \bibinfo {pages} {090503}
  (\bibinfo {year} {2005})}\BibitemShut {NoStop}%
\bibitem [{\citenamefont {Or{\'u}s}(2014)}]{orus2014practical}%
  \BibitemOpen
  \bibfield  {author} {\bibinfo {author} {\bibfnamefont {R.}~\bibnamefont
  {Or{\'u}s}},\ }\href@noop {} {\bibfield  {journal} {\bibinfo  {journal}
  {Annals of Physics}\ }\textbf {\bibinfo {volume} {349}},\ \bibinfo {pages}
  {117} (\bibinfo {year} {2014})}\BibitemShut {NoStop}%
\bibitem [{\citenamefont {Bridgeman}\ and\ \citenamefont
  {Chubb}(2017)}]{bridgeman2017hand}%
  \BibitemOpen
  \bibfield  {author} {\bibinfo {author} {\bibfnamefont {J.~C.}\ \bibnamefont
  {Bridgeman}}\ and\ \bibinfo {author} {\bibfnamefont {C.~T.}\ \bibnamefont
  {Chubb}},\ }\href@noop {} {\bibfield  {journal} {\bibinfo  {journal} {J.
  Phys. A}\ }\textbf {\bibinfo {volume} {50}},\ \bibinfo {pages} {223001}
  (\bibinfo {year} {2017})}\BibitemShut {NoStop}%
\bibitem [{\citenamefont {Lanczos}(1950)}]{lanczos1950iteration}%
  \BibitemOpen
  \bibfield  {author} {\bibinfo {author} {\bibfnamefont {C.}~\bibnamefont
  {Lanczos}},\ }\href@noop {} {\emph {\bibinfo {title} {An iteration method for
  the solution of the eigenvalue problem of linear differential and integral
  operators}}}\ (\bibinfo  {publisher} {United States Governm. Press Office Los
  Angeles, CA},\ \bibinfo {year} {1950})\BibitemShut {NoStop}%
\bibitem [{\citenamefont {Golub}\ and\ \citenamefont
  {Meurant}(1994)}]{golub1994matrices}%
  \BibitemOpen
  \bibfield  {author} {\bibinfo {author} {\bibfnamefont {G.~H.}\ \bibnamefont
  {Golub}}\ and\ \bibinfo {author} {\bibfnamefont {G.}~\bibnamefont
  {Meurant}},\ }\href@noop {} {\bibfield  {journal} {\bibinfo  {journal}
  {Pitman Research Notes in Mathematics Series}\ ,\ \bibinfo {pages} {105}}
  (\bibinfo {year} {1994})}\BibitemShut {NoStop}%
\bibitem [{\citenamefont {Ubaru}\ \emph {et~al.}(2017)\citenamefont {Ubaru},
  \citenamefont {Chen},\ and\ \citenamefont {Saad}}]{ubaru2017fast}%
  \BibitemOpen
  \bibfield  {author} {\bibinfo {author} {\bibfnamefont {S.}~\bibnamefont
  {Ubaru}}, \bibinfo {author} {\bibfnamefont {J.}~\bibnamefont {Chen}}, \ and\
  \bibinfo {author} {\bibfnamefont {Y.}~\bibnamefont {Saad}},\ }\href@noop {}
  {\bibfield  {journal} {\bibinfo  {journal} {SIAM Journal on Matrix Analysis
  and Applications}\ }\textbf {\bibinfo {volume} {38}},\ \bibinfo {pages}
  {1075} (\bibinfo {year} {2017})}\BibitemShut {NoStop}%
\bibitem [{\citenamefont {Gray}(2018)}]{gray2018quimb}%
  \BibitemOpen
  \bibfield  {author} {\bibinfo {author} {\bibfnamefont {J.}~\bibnamefont
  {Gray}},\ }\href {\doibase 10.21105/joss.00819} {\bibfield  {journal}
  {\bibinfo  {journal} {Journal of Open Source Software}\ }\textbf {\bibinfo
  {volume} {3}},\ \bibinfo {pages} {819} (\bibinfo {year} {2018})}\BibitemShut
  {NoStop}%
\bibitem [{\citenamefont {Peres}(1996)}]{peres1996separability}%
  \BibitemOpen
  \bibfield  {author} {\bibinfo {author} {\bibfnamefont {A.}~\bibnamefont
  {Peres}},\ }\href@noop {} {\bibfield  {journal} {\bibinfo  {journal}
  {Physical Review Letters}\ }\textbf {\bibinfo {volume} {77}},\ \bibinfo
  {pages} {1413} (\bibinfo {year} {1996})}\BibitemShut {NoStop}%
\bibitem [{\citenamefont {Han}\ \emph {et~al.}(2016)\citenamefont {Han},
  \citenamefont {Malioutov}, \citenamefont {Avron},\ and\ \citenamefont
  {Shin}}]{han2016approximating}%
  \BibitemOpen
  \bibfield  {author} {\bibinfo {author} {\bibfnamefont {I.}~\bibnamefont
  {Han}}, \bibinfo {author} {\bibfnamefont {D.}~\bibnamefont {Malioutov}},
  \bibinfo {author} {\bibfnamefont {H.}~\bibnamefont {Avron}}, \ and\ \bibinfo
  {author} {\bibfnamefont {J.}~\bibnamefont {Shin}},\ }\href@noop {} {\bibfield
   {journal} {\bibinfo  {journal} {arXiv:1606.00942}\ } (\bibinfo {year}
  {2016})}\BibitemShut {NoStop}%
\bibitem [{\citenamefont {Lin}\ \emph {et~al.}(2016)\citenamefont {Lin},
  \citenamefont {Saad},\ and\ \citenamefont {Yang}}]{lin2016approximating}%
  \BibitemOpen
  \bibfield  {author} {\bibinfo {author} {\bibfnamefont {L.}~\bibnamefont
  {Lin}}, \bibinfo {author} {\bibfnamefont {Y.}~\bibnamefont {Saad}}, \ and\
  \bibinfo {author} {\bibfnamefont {C.}~\bibnamefont {Yang}},\ }\href@noop {}
  {\bibfield  {journal} {\bibinfo  {journal} {SIAM review}\ }\textbf {\bibinfo
  {volume} {58}},\ \bibinfo {pages} {34} (\bibinfo {year} {2016})}\BibitemShut
  {NoStop}%
\bibitem [{\citenamefont {Hutchinson}(1990)}]{hutchinson1990stochastic}%
  \BibitemOpen
  \bibfield  {author} {\bibinfo {author} {\bibfnamefont {M.~F.}\ \bibnamefont
  {Hutchinson}},\ }\href@noop {} {\bibfield  {journal} {\bibinfo  {journal}
  {Commun. Stat. Simul. Comput.}\ }\textbf {\bibinfo {volume} {19}},\ \bibinfo
  {pages} {433} (\bibinfo {year} {1990})}\BibitemShut {NoStop}%
\bibitem [{\citenamefont {Bellalij}\ \emph {et~al.}(2015)\citenamefont
  {Bellalij}, \citenamefont {Reichel}, \citenamefont {Rodriguez},\ and\
  \citenamefont {Sadok}}]{bellalij2015bounding}%
  \BibitemOpen
  \bibfield  {author} {\bibinfo {author} {\bibfnamefont {M.}~\bibnamefont
  {Bellalij}}, \bibinfo {author} {\bibfnamefont {L.}~\bibnamefont {Reichel}},
  \bibinfo {author} {\bibfnamefont {G.}~\bibnamefont {Rodriguez}}, \ and\
  \bibinfo {author} {\bibfnamefont {H.}~\bibnamefont {Sadok}},\ }\href@noop {}
  {\bibfield  {journal} {\bibinfo  {journal} {Applied Numerical Mathematics}\
  }\textbf {\bibinfo {volume} {94}},\ \bibinfo {pages} {127} (\bibinfo {year}
  {2015})}\BibitemShut {NoStop}%
\bibitem [{\citenamefont {Biamonte}\ and\ \citenamefont
  {Bergholm}(2017)}]{biamonte2017tensor}%
  \BibitemOpen
  \bibfield  {author} {\bibinfo {author} {\bibfnamefont {J.}~\bibnamefont
  {Biamonte}}\ and\ \bibinfo {author} {\bibfnamefont {V.}~\bibnamefont
  {Bergholm}},\ }\href@noop {} {\bibfield  {journal} {\bibinfo  {journal}
  {arXiv preprint arXiv:1708.00006}\ } (\bibinfo {year} {2017})}\BibitemShut
  {NoStop}%
\bibitem [{\citenamefont {Liberty}\ \emph {et~al.}(2007)\citenamefont
  {Liberty}, \citenamefont {Woolfe}, \citenamefont {Martinsson}, \citenamefont
  {Rokhlin},\ and\ \citenamefont {Tygert}}]{liberty2007randomized}%
  \BibitemOpen
  \bibfield  {author} {\bibinfo {author} {\bibfnamefont {E.}~\bibnamefont
  {Liberty}}, \bibinfo {author} {\bibfnamefont {F.}~\bibnamefont {Woolfe}},
  \bibinfo {author} {\bibfnamefont {P.-G.}\ \bibnamefont {Martinsson}},
  \bibinfo {author} {\bibfnamefont {V.}~\bibnamefont {Rokhlin}}, \ and\
  \bibinfo {author} {\bibfnamefont {M.}~\bibnamefont {Tygert}},\ }\href@noop {}
  {\bibfield  {journal} {\bibinfo  {journal} {Proceedings of the National
  Academy of Sciences}\ }\textbf {\bibinfo {volume} {104}},\ \bibinfo {pages}
  {20167} (\bibinfo {year} {2007})}\BibitemShut {NoStop}%
\bibitem [{\citenamefont {Woolfe}\ \emph {et~al.}(2008)\citenamefont {Woolfe},
  \citenamefont {Liberty}, \citenamefont {Rokhlin},\ and\ \citenamefont
  {Tygert}}]{woolfe2008fast}%
  \BibitemOpen
  \bibfield  {author} {\bibinfo {author} {\bibfnamefont {F.}~\bibnamefont
  {Woolfe}}, \bibinfo {author} {\bibfnamefont {E.}~\bibnamefont {Liberty}},
  \bibinfo {author} {\bibfnamefont {V.}~\bibnamefont {Rokhlin}}, \ and\
  \bibinfo {author} {\bibfnamefont {M.}~\bibnamefont {Tygert}},\ }\href@noop {}
  {\bibfield  {journal} {\bibinfo  {journal} {Applied and Computational
  Harmonic Analysis}\ }\textbf {\bibinfo {volume} {25}},\ \bibinfo {pages}
  {335} (\bibinfo {year} {2008})}\BibitemShut {NoStop}%
\bibitem [{\citenamefont {Martinsson}\ \emph {et~al.}(2011)\citenamefont
  {Martinsson}, \citenamefont {Rokhlin},\ and\ \citenamefont
  {Tygert}}]{martinsson2011randomized}%
  \BibitemOpen
  \bibfield  {author} {\bibinfo {author} {\bibfnamefont {P.-G.}\ \bibnamefont
  {Martinsson}}, \bibinfo {author} {\bibfnamefont {V.}~\bibnamefont {Rokhlin}},
  \ and\ \bibinfo {author} {\bibfnamefont {M.}~\bibnamefont {Tygert}},\
  }\href@noop {} {\bibfield  {journal} {\bibinfo  {journal} {Applied and
  Computational Harmonic Analysis}\ }\textbf {\bibinfo {volume} {30}},\
  \bibinfo {pages} {47} (\bibinfo {year} {2011})}\BibitemShut {NoStop}%
\bibitem [{\citenamefont {Hastings}(2007)}]{hastings2007area}%
  \BibitemOpen
  \bibfield  {author} {\bibinfo {author} {\bibfnamefont {M.~B.}\ \bibnamefont
  {Hastings}},\ }\href@noop {} {\bibfield  {journal} {\bibinfo  {journal}
  {Journal of Statistical Mechanics: Theory and Experiment}\ }\textbf {\bibinfo
  {volume} {2007}},\ \bibinfo {pages} {P08024} (\bibinfo {year}
  {2007})}\BibitemShut {NoStop}%
\bibitem [{\citenamefont {White}(1992)}]{white1992density}%
  \BibitemOpen
  \bibfield  {author} {\bibinfo {author} {\bibfnamefont {S.~R.}\ \bibnamefont
  {White}},\ }\href@noop {} {\bibfield  {journal} {\bibinfo  {journal}
  {Physical review letters}\ }\textbf {\bibinfo {volume} {69}},\ \bibinfo
  {pages} {2863} (\bibinfo {year} {1992})}\BibitemShut {NoStop}%
\bibitem [{\citenamefont {Vidal}(2003)}]{vidal2003efficient}%
  \BibitemOpen
  \bibfield  {author} {\bibinfo {author} {\bibfnamefont {G.}~\bibnamefont
  {Vidal}},\ }\href@noop {} {\bibfield  {journal} {\bibinfo  {journal} {Phys.
  Rev. Lett.}\ }\textbf {\bibinfo {volume} {91}},\ \bibinfo {pages} {147902}
  (\bibinfo {year} {2003})}\BibitemShut {NoStop}%
\bibitem [{\citenamefont {Pippan}\ \emph {et~al.}(2010)\citenamefont {Pippan},
  \citenamefont {White},\ and\ \citenamefont {Evertz}}]{pippan2010efficient}%
  \BibitemOpen
  \bibfield  {author} {\bibinfo {author} {\bibfnamefont {P.}~\bibnamefont
  {Pippan}}, \bibinfo {author} {\bibfnamefont {S.~R.}\ \bibnamefont {White}}, \
  and\ \bibinfo {author} {\bibfnamefont {H.~G.}\ \bibnamefont {Evertz}},\
  }\href@noop {} {\bibfield  {journal} {\bibinfo  {journal} {Phys. Rev. B}\
  }\textbf {\bibinfo {volume} {81}},\ \bibinfo {pages} {081103} (\bibinfo
  {year} {2010})}\BibitemShut {NoStop}%
\bibitem [{\citenamefont {Ruggiero}\ \emph {et~al.}(2016)\citenamefont
  {Ruggiero}, \citenamefont {Alba},\ and\ \citenamefont
  {Calabrese}}]{ruggiero2016entanglement}%
  \BibitemOpen
  \bibfield  {author} {\bibinfo {author} {\bibfnamefont {P.}~\bibnamefont
  {Ruggiero}}, \bibinfo {author} {\bibfnamefont {V.}~\bibnamefont {Alba}}, \
  and\ \bibinfo {author} {\bibfnamefont {P.}~\bibnamefont {Calabrese}},\
  }\href@noop {} {\bibfield  {journal} {\bibinfo  {journal} {Physical Review
  B}\ }\textbf {\bibinfo {volume} {94}},\ \bibinfo {pages} {035152} (\bibinfo
  {year} {2016})}\BibitemShut {NoStop}%
\bibitem [{\citenamefont {Mbeng}\ \emph {et~al.}(2017)\citenamefont {Mbeng},
  \citenamefont {Alba},\ and\ \citenamefont {Calabrese}}]{mbeng2017negativity}%
  \BibitemOpen
  \bibfield  {author} {\bibinfo {author} {\bibfnamefont {G.~B.}\ \bibnamefont
  {Mbeng}}, \bibinfo {author} {\bibfnamefont {V.}~\bibnamefont {Alba}}, \ and\
  \bibinfo {author} {\bibfnamefont {P.}~\bibnamefont {Calabrese}},\ }\href@noop
  {} {\bibfield  {journal} {\bibinfo  {journal} {Journal of Physics A:
  Mathematical and Theoretical}\ }\textbf {\bibinfo {volume} {50}},\ \bibinfo
  {pages} {194001} (\bibinfo {year} {2017})}\BibitemShut {NoStop}%
\bibitem [{\citenamefont {Smith}\ and\ \citenamefont
  {Gray}(2018)}]{smith2018opteinsum}%
  \BibitemOpen
  \bibfield  {author} {\bibinfo {author} {\bibfnamefont {D.~G.}\ \bibnamefont
  {Smith}}\ and\ \bibinfo {author} {\bibfnamefont {J.}~\bibnamefont {Gray}},\
  }\href {\doibase 10.21105/joss.00753} {\bibfield  {journal} {\bibinfo
  {journal} {Journal of Open Source Software}\ }\textbf {\bibinfo {volume}
  {3}},\ \bibinfo {pages} {753} (\bibinfo {year} {2018})}\BibitemShut {NoStop}%
\bibitem [{\citenamefont {Bhosale}\ \emph {et~al.}(2012)\citenamefont
  {Bhosale}, \citenamefont {Tomsovic},\ and\ \citenamefont
  {Lakshminarayan}}]{bhosale2012entanglement}%
  \BibitemOpen
  \bibfield  {author} {\bibinfo {author} {\bibfnamefont {U.~T.}\ \bibnamefont
  {Bhosale}}, \bibinfo {author} {\bibfnamefont {S.}~\bibnamefont {Tomsovic}}, \
  and\ \bibinfo {author} {\bibfnamefont {A.}~\bibnamefont {Lakshminarayan}},\
  }\href@noop {} {\bibfield  {journal} {\bibinfo  {journal} {Physical Review
  A}\ }\textbf {\bibinfo {volume} {85}},\ \bibinfo {pages} {062331} (\bibinfo
  {year} {2012})}\BibitemShut {NoStop}%
\bibitem [{\citenamefont {Sekino}\ and\ \citenamefont
  {Susskind}(2008)}]{sekino2008fast}%
  \BibitemOpen
  \bibfield  {author} {\bibinfo {author} {\bibfnamefont {Y.}~\bibnamefont
  {Sekino}}\ and\ \bibinfo {author} {\bibfnamefont {L.}~\bibnamefont
  {Susskind}},\ }\href@noop {} {\bibfield  {journal} {\bibinfo  {journal}
  {Journal of High Energy Physics}\ }\textbf {\bibinfo {volume} {2008}},\
  \bibinfo {pages} {065} (\bibinfo {year} {2008})}\BibitemShut {NoStop}%
\bibitem [{\citenamefont {Avron}\ and\ \citenamefont
  {Toledo}(2011)}]{avron2011randomized}%
  \BibitemOpen
  \bibfield  {author} {\bibinfo {author} {\bibfnamefont {H.}~\bibnamefont
  {Avron}}\ and\ \bibinfo {author} {\bibfnamefont {S.}~\bibnamefont {Toledo}},\
  }\href@noop {} {\bibfield  {journal} {\bibinfo  {journal} {Journal of the ACM
  (JACM)}\ }\textbf {\bibinfo {volume} {58}},\ \bibinfo {pages} {8} (\bibinfo
  {year} {2011})}\BibitemShut {NoStop}%
\bibitem [{\citenamefont {August}\ \emph {et~al.}(2017)\citenamefont {August},
  \citenamefont {Ba{\~n}uls},\ and\ \citenamefont
  {Huckle}}]{august2017approximation}%
  \BibitemOpen
  \bibfield  {author} {\bibinfo {author} {\bibfnamefont {M.}~\bibnamefont
  {August}}, \bibinfo {author} {\bibfnamefont {M.~C.}\ \bibnamefont
  {Ba{\~n}uls}}, \ and\ \bibinfo {author} {\bibfnamefont {T.}~\bibnamefont
  {Huckle}},\ }\href@noop {} {\bibfield  {journal} {\bibinfo  {journal}
  {Electronic Transactions on Numerical Analysis}\ }\textbf {\bibinfo {volume}
  {46}},\ \bibinfo {pages} {215} (\bibinfo {year} {2017})}\BibitemShut
  {NoStop}%
\bibitem [{\citenamefont {Shi}\ \emph {et~al.}(2006)\citenamefont {Shi},
  \citenamefont {Duan},\ and\ \citenamefont {Vidal}}]{shi2006classical}%
  \BibitemOpen
  \bibfield  {author} {\bibinfo {author} {\bibfnamefont {Y.-Y.}\ \bibnamefont
  {Shi}}, \bibinfo {author} {\bibfnamefont {L.-M.}\ \bibnamefont {Duan}}, \
  and\ \bibinfo {author} {\bibfnamefont {G.}~\bibnamefont {Vidal}},\
  }\href@noop {} {\bibfield  {journal} {\bibinfo  {journal} {Physical review
  a}\ }\textbf {\bibinfo {volume} {74}},\ \bibinfo {pages} {022320} (\bibinfo
  {year} {2006})}\BibitemShut {NoStop}%
\bibitem [{\citenamefont {Murg}\ \emph {et~al.}(2010)\citenamefont {Murg},
  \citenamefont {Verstraete}, \citenamefont {Legeza},\ and\ \citenamefont
  {Noack}}]{murg2010simulating}%
  \BibitemOpen
  \bibfield  {author} {\bibinfo {author} {\bibfnamefont {V.}~\bibnamefont
  {Murg}}, \bibinfo {author} {\bibfnamefont {F.}~\bibnamefont {Verstraete}},
  \bibinfo {author} {\bibfnamefont {{\"O}.}~\bibnamefont {Legeza}}, \ and\
  \bibinfo {author} {\bibfnamefont {R.-h.~M.}\ \bibnamefont {Noack}},\
  }\href@noop {} {\bibfield  {journal} {\bibinfo  {journal} {Physical Review
  B}\ }\textbf {\bibinfo {volume} {82}},\ \bibinfo {pages} {205105} (\bibinfo
  {year} {2010})}\BibitemShut {NoStop}%
\bibitem [{\citenamefont {Vidal}(2007)}]{vidal2007entanglement}%
  \BibitemOpen
  \bibfield  {author} {\bibinfo {author} {\bibfnamefont {G.}~\bibnamefont
  {Vidal}},\ }\href@noop {} {\bibfield  {journal} {\bibinfo  {journal}
  {Physical review letters}\ }\textbf {\bibinfo {volume} {99}},\ \bibinfo
  {pages} {220405} (\bibinfo {year} {2007})}\BibitemShut {NoStop}%
\bibitem [{\citenamefont {Vidal}(2008)}]{vidal2008class}%
  \BibitemOpen
  \bibfield  {author} {\bibinfo {author} {\bibfnamefont {G.}~\bibnamefont
  {Vidal}},\ }\href@noop {} {\bibfield  {journal} {\bibinfo  {journal}
  {Physical review letters}\ }\textbf {\bibinfo {volume} {101}},\ \bibinfo
  {pages} {110501} (\bibinfo {year} {2008})}\BibitemShut {NoStop}%
\bibitem [{\citenamefont {Markov}\ and\ \citenamefont
  {Shi}(2008)}]{markov2008simulating}%
  \BibitemOpen
  \bibfield  {author} {\bibinfo {author} {\bibfnamefont {I.~L.}\ \bibnamefont
  {Markov}}\ and\ \bibinfo {author} {\bibfnamefont {Y.}~\bibnamefont {Shi}},\
  }\href@noop {} {\bibfield  {journal} {\bibinfo  {journal} {SIAM Journal on
  Computing}\ }\textbf {\bibinfo {volume} {38}},\ \bibinfo {pages} {963}
  (\bibinfo {year} {2008})}\BibitemShut {NoStop}%
\bibitem [{\citenamefont {Pednault}\ \emph {et~al.}(2017)\citenamefont
  {Pednault}, \citenamefont {Gunnels}, \citenamefont {Nannicini}, \citenamefont
  {Horesh}, \citenamefont {Magerlein}, \citenamefont {Solomonik},\ and\
  \citenamefont {Wisnieff}}]{pednault2017breaking}%
  \BibitemOpen
  \bibfield  {author} {\bibinfo {author} {\bibfnamefont {E.}~\bibnamefont
  {Pednault}}, \bibinfo {author} {\bibfnamefont {J.~A.}\ \bibnamefont
  {Gunnels}}, \bibinfo {author} {\bibfnamefont {G.}~\bibnamefont {Nannicini}},
  \bibinfo {author} {\bibfnamefont {L.}~\bibnamefont {Horesh}}, \bibinfo
  {author} {\bibfnamefont {T.}~\bibnamefont {Magerlein}}, \bibinfo {author}
  {\bibfnamefont {E.}~\bibnamefont {Solomonik}}, \ and\ \bibinfo {author}
  {\bibfnamefont {R.}~\bibnamefont {Wisnieff}},\ }\href@noop {} {\bibfield
  {journal} {\bibinfo  {journal} {arXiv preprint arXiv:1710.05867}\ } (\bibinfo
  {year} {2017})}\BibitemShut {NoStop}%
\bibitem [{\citenamefont {Boixo}\ \emph {et~al.}(2017)\citenamefont {Boixo},
  \citenamefont {Isakov}, \citenamefont {Smelyanskiy},\ and\ \citenamefont
  {Neven}}]{boixo2017simulation}%
  \BibitemOpen
  \bibfield  {author} {\bibinfo {author} {\bibfnamefont {S.}~\bibnamefont
  {Boixo}}, \bibinfo {author} {\bibfnamefont {S.~V.}\ \bibnamefont {Isakov}},
  \bibinfo {author} {\bibfnamefont {V.~N.}\ \bibnamefont {Smelyanskiy}}, \ and\
  \bibinfo {author} {\bibfnamefont {H.}~\bibnamefont {Neven}},\ }\href@noop {}
  {\bibfield  {journal} {\bibinfo  {journal} {arXiv preprint arXiv:1712.05384}\
  } (\bibinfo {year} {2017})}\BibitemShut {NoStop}%
\bibitem [{\citenamefont {Chen}\ \emph {et~al.}(2018)\citenamefont {Chen},
  \citenamefont {Zhang}, \citenamefont {Chen}, \citenamefont {Huang},
  \citenamefont {Newman},\ and\ \citenamefont {Shi}}]{chen2018classical}%
  \BibitemOpen
  \bibfield  {author} {\bibinfo {author} {\bibfnamefont {J.}~\bibnamefont
  {Chen}}, \bibinfo {author} {\bibfnamefont {F.}~\bibnamefont {Zhang}},
  \bibinfo {author} {\bibfnamefont {M.}~\bibnamefont {Chen}}, \bibinfo {author}
  {\bibfnamefont {C.}~\bibnamefont {Huang}}, \bibinfo {author} {\bibfnamefont
  {M.}~\bibnamefont {Newman}}, \ and\ \bibinfo {author} {\bibfnamefont
  {Y.}~\bibnamefont {Shi}},\ }\href@noop {} {\bibfield  {journal} {\bibinfo
  {journal} {arXiv preprint arXiv:1805.01450}\ } (\bibinfo {year}
  {2018})}\BibitemShut {NoStop}%
\bibitem [{\citenamefont {Markov}\ \emph {et~al.}(2018)\citenamefont {Markov},
  \citenamefont {Fatima}, \citenamefont {Isakov},\ and\ \citenamefont
  {Boixo}}]{markov2018quantum}%
  \BibitemOpen
  \bibfield  {author} {\bibinfo {author} {\bibfnamefont {I.~L.}\ \bibnamefont
  {Markov}}, \bibinfo {author} {\bibfnamefont {A.}~\bibnamefont {Fatima}},
  \bibinfo {author} {\bibfnamefont {S.~V.}\ \bibnamefont {Isakov}}, \ and\
  \bibinfo {author} {\bibfnamefont {S.}~\bibnamefont {Boixo}},\ }\href@noop {}
  {\bibfield  {journal} {\bibinfo  {journal} {arXiv preprint arXiv:1807.10749}\
  } (\bibinfo {year} {2018})}\BibitemShut {NoStop}%
\end{thebibliography}%

\end{document}